\documentclass[12pt, draftclsnofoot, onecolumn]{IEEEtran}

\usepackage{amsfonts}
\usepackage{float}
\usepackage{amssymb}
\usepackage[T1]{fontenc}
\usepackage{amsthm}
\usepackage{units}
\usepackage{multicol}
\usepackage{multirow}
\usepackage{color}

\newtheorem{lem}{Lemma}

\newtheorem{prop}{Proposition}

\usepackage{cite}
\ifCLASSINFOpdf
  \usepackage[pdftex]{graphicx}
   \graphicspath{{../pdf/}{../jpeg/}}
  \DeclareGraphicsExtensions{.pdf,.jpeg,.png}
\else
  \usepackage[dvips]{graphicx}
  \graphicspath{{../eps/}}
  \DeclareGraphicsExtensions{.eps}
\fi
\usepackage[cmex10]{amsmath}

\begin{document}
\title{Distributed User Scheduling for MIMO-Y Channel}
\author{Hui~Gao,~\emph{Member},~\emph{IEEE},{~Chau~Yuen},~\emph{Senior Member},~\emph{IEEE},~{Yuan~Ren},~{Wei~Long},~{Tiejun~Lv},~\emph{Senior Member},~\emph{IEEE}

%       \thanks{Manuscript received December 26, 2011; revised June 6, 2012 and September 14, 2012; accepted September 24, 2012. T. Lv is the corresponding author. The work of H. Gao was done during his Ph.D. study at Beijing University of Posts and Telecommunications. The work of H. Gao and T. Lv was supported in part by NSFC (Grant No. 60972075 and No. 61271188). The work of S. Zhang was supported in part by NSFC (Grant No. 20902016) and NSF Guangdong (Grant No. 10151806001000003). The work of C. Yuen was partly supported by the Singapore University Technology and Design (Grant No. SUTD-ZJU/RES/02/2011). The work of S. Yang was partly supported by China Scholarship Council. This paper was presented in part at the 2012 IEEE International Conference on Communications, Ottawa, Canada. The review of this paper was coordinated by Prof. Masoud Ardakani.}

\thanks{H. Gao, R. Yuan, W. Long and T. Lv are with the School of Information and Communication Engineering, Beijing University of Posts and Telecommunications, Beijing, 100876, China (e-mail: huigao@bupt.edu.cn, renyuan@bupt.edu.cn, longwei@bupt.edu.cn, lvtiejun@bupt.edu.cn).}
\thanks{C. Yuen are with Singapore University of Technology and Design, Singapore, 138682. (e-mail: yuenchau@sutd.edu.sg).}}

%%\thanks{S. Yang is with the School of Electronics and
%%Computer Science, University of Southampton, SO17 1BJ Southampton, U.K. (e-mail: sy7g09@ecs.soton.ac.uk).
%}

% <-this % stops a space
%\thanks{Manuscript received April 19, 2005; revised January 11, 2007.}

%\markboth{THE WORK IS SUBMITTED TO IEEE TRANSACTIONS ON WIRELESS COMMUNICATIONS, August, 2014}%
{}



\maketitle
\begin{abstract}
In this paper, distributed user scheduling schemes are proposed for
the multi-user MIMO-Y channel, where three $N_{T}$-antenna users
(\textcolor{black}{$N_{T}=2N,\,3N$}) are selected from three clusters
to exchange information via an \textcolor{black}{$N_{R}$}-antenna
amplify-and-forward (AF) relay\textcolor{black}{{} ($N_{R}=3N$), and
$N\geq1$ represents the number of data stream(s) of each unicast
transmission within the MIMO-Y channel}. The proposed schemes effectively
harvest multi-user diversity (MuD) without the need of global channel
state information (CSI) or centralized computations. In particular,
a novel reference signal space (RSS) is proposed to enable the distributed
scheduling for both cluster-wise (CS) and group-wise (GS) patterns.
The minimum user-antenna (Min-UA) transmission with\textcolor{black}{{}
$N_{T}=2N$ }is first considered. Next, we consider an equal number
of relay and user antenna (ER-UA) transmission with ${\color{black}N_{T}=3N}$,
with the aim of reducing CSI overhead as compared to Min-UA. For ER-UA
transmission, the achievable MuD orders of the proposed distributed
scheduling schemes are analytically derived, which proves the superiority
and optimality of the proposed RSS-based distributed scheduling. These
results reveal some fundamental behaviors of MuD and the performance-complexity
tradeoff of user scheduling schemes in the MIMO-Y channel. \end{abstract}

% \fi
\IEEEpeerreviewmaketitle

\section{Introduction}
\IEEEPARstart{}{T}{he} {\color{black}multi-way relay channel}
\cite{Multiway_Relay} has been considered as a fundamental building
block for future cooperative communications. {\color{black}The capacity
and degrees-of-freedom of multi-way relay channel have been partially
studied by \cite{chaaban2014channel,ong2011capacity,cui2012communication}
assuming specific system configurations and traffic patterns. The
more general multi-group multi-way relay scenario is considered in
\cite{degenhardt2013non} with a novel and efficient joint spatial and
temporal signal processing.} Currently, two basic representatives
of the {\color{black}multi-way relay channel} have been focused.
The first one is the two-way relay channel (TWRC), which has been
extensively studied with various wireless network coding (WNC) techniques
\cite{SigANC,PNC,ling2010differential}. {\color{black} Recent information-theoretic studies
on TWRC can be found in \cite{avestimehr2010capacity,sezgin2012divide}
and references therein. In particular, the capacity
region of TWRC is characterized in \cite{avestimehr2010capacity}
with the deterministic approach, then the linear shift deterministic
model is employed in \cite{sezgin2012divide} to analyze the capacity
region of the multi-pair TWRC.} The second representative is the MIMO-Y channel
\cite{lee2010degrees}, which is a novel extension of the TWRC with
multiple independent unicast transmissions among three users. As compared
to the TWRC, the MIMO-Y channel requires more sophisticated signal
processing with WNC and spatial-resource management. Specifically,
the basic MIMO-Y channel has been proposed with a new concept of signal
space alignment (SSA) \cite{lee2010degrees}, which is a novel application
of the principle of interference alignment \cite{cadambe2008interference}.
SSA aligns the bi-directional information from two users at the relay
to maximize the utility of the relay antennas, and it also enables
the WNC \cite{SigANC,PNC} for efficient transmission with the half-duplex
relay. Because of its fundamental role and novel transmission schemes,
the MIMO-Y channel is now attracting increasing attentions.

The MIMO-Y channel can find many interesting applications in various
three-party communication scenarios. For example, in ad-hoc networks,
three geographically isolated nodes can exchange messages with the
help of the relay; in cellular networks, a group of three users can
share information via the relay with flexible cooperative or device-to-device
communication protocols; in satellite communication, the satellite
often serves as a relay to enable information exchanges among three
earth stations. Inspired by the wide range of potential applications,
many efforts have been devoted to understanding the fundamental limit
of MIMO-Y channel. For example, the achievable degrees-of-freedom (DoF)
and capacity of the MIMO-Y channel have been studied with various
antenna configurations at the users or relay \cite{lee2010degrees,MIMO-Y-K,chaaban2013approximate,chaaban14approximate,LongICC,yuan2014achievable}.
{\color{black}Specifically, the original SSA scheme in \cite{lee2010degrees}
has been extended to the generalized $K$-user MIMO-Y channel in \cite{MIMO-Y-K}.
For the single antenna Gaussian Y-channel, the approximate sum-capacity
and the capacity region are characterized in \cite{chaaban2013approximate}
and \cite{chaaban14approximate} separately. Recently, an asymmetric
SSA scheme is proposed in \cite{LongICC} to assist
the information exchange of single antenna users and the achievable
DoF of the four-user MIMO-Y channel is investigated in
\cite{yuan2014achievable}.} Beyond the concerns of the fundamental
limit, some practical schemes have been also proposed to enhance the
transmission reliability of MIMO-Y channel in the wireless fading
environment \cite{ding2011generalized,ZhendongICC,Teav}. Of particular
interests are the diversity-achieving beamforming schemes, which employ
\textcolor{black}{extra antennas at the user, i.e., more antennas
than the minimum requirement of SSA operation,} to perform selective or iterative
beamforming optimization \cite{ding2011generalized,ZhendongICC}.
Although these schemes show significant performance improvements as
compared to the proof-of-concept scheme in \cite{lee2010degrees},
solely relying on the user's \textcolor{black}{redundant} antennas
for a scalable diversity gain is not always practical. The limited
size and power supply of the user's equipment are practical constraints.
Therefore, other diversity-achieving schemes are also demanded to
complement these beamforming techniques.

Multi-user diversity (MuD), which is known as an important source
to combat wireless fading \cite{FW,GFeedback}, can be potentially
exploited for the MIMO-Y channel. It has been noted that although
the number of antennas is limited for each user's equipment, a system
potentially has multiple users requiring data transmission. Therefore,
by carefully scheduling the users' transmissions, significant performance
gain can be obtained. The multi-user scheduling has been studied for
the traditional broadcasting \cite{Taesang} and {\color{black}multi-user}
interference channels \cite{Yang_OIA,Jung_OIA}. Regarding the general
{\color{black}multi-way relay channel}, the comprehensive solution
of user scheduling is still open. Some initial researches have been
done for the TWRC with a variety of system configurations \cite{upadhyay2011outage,Jeon2012opp,Joung2012user,Wang2012,Jang2010P,Xiang2012},
and they offer valuable insights to inspire new applications. However,
the designs of efficient user scheduling schemes for the MIMO-Y channel
have different challenges as compared to {\color{black}the} TWRC. In general,
the MIMO-Y channel calls for new user scheduling methods for its unique
system and traffic configurations, i.e., each user has multiple antennas
to support two independent unicast information flows \cite{lee2010degrees}.
In particular, the unique SSA-oriented MIMO-Y transmission requires
more sophisticated transmit/receive beamforming designs\cite{lee2010degrees},
which are often coupled with the multi-user scheduling metrics \cite{GaoICC}.
Such coupling may significantly increases the system overheads for
CSI and the computation complexity of the scheduling center. Taking
the scheduling methods \cite{Taesang,upadhyay2011outage,Jeon2012opp,Joung2012user,Wang2012,Jang2010P,Xiang2012,GaoICC}
for example, they are all conducted in a centralized fashion with
global CSI and require relatively complicated computations at the
scheduling center. In fact, even in the cellular network with high
user density, asking the base station to learn the global CSI is costly
\cite{GFeedback}. For the MIMO-Y channel, which often fits into the
low-complexity and structure-less networks, the assumption of a powerful
dedicated scheduling center is not always feasible, especially when
one node or user just serves as the immediate relay. {\color{black}Therefore,
novel cost-effective scheduling methods are needed for the MIMO-Y
channel. As an initial study on this issue, a distributed scheduling scheme with sketchy performance
analysis is reported in \cite{HuiGCworkshop}.}

%In this paper, we consider... move the green color paragraph here... In particular, we propose a low-complexity ...... with two scheduling patterns. The two scheduling patterns are cluster-wise shceduling (CS) and group-wise scheduling (GS).  For the CS......

In this paper, we consider a basic multi-user MIMO-Y channel, where
one $N_{R}$-antenna relay \textcolor{black}{($N_{R}=3N$)} helps information
exchange among three selected $N_{T}$-antenna (\textcolor{black}{$N_{T}=2N,\,3N$})
users from three clusters\textcolor{black}{, and $N\geq1$ represents
the number of data stream(s) of each unicast transmission within the
MIMO-Y channel}. Such basic configuration is sufficient to capture
the essential of the MIMO-Y transmission; it also simplifies MuD analysis
for clear insight. In particular, we propose low-complexity distributed
user scheduling schemes for the MIMO-Y channel with two scheduling
patterns, namely, cluster-wise scheduling (CS) and group-wise scheduling
(GS). For the CS, a cluster representative is selected from each cluster,
and the three selected representatives conduct information exchange
via the relay. Such scheduling may find applications in the wireless
ad-hoc or sensor networks. For example, when some globally critical
events are observed by many on-site nodes at three isolated places,
one node is selected from each cluster to perform information exchange.
For the GS, three users (each from a different cluster) are associated
within a predefined group before transmission, and one group is scheduled
to exchange information via the relay. Such scheduling may be useful
in the cellular networks or device-to-device networks where a group
of three users {\color{black}wishes} to share information within
their social network.

Moreover, depending on the number of required antennas equipped at
the user, two possible MIMO-Y transmission schemes are considered.
Specifically, the transmission scheme with the \emph{minimum} number
of \emph{user} \emph{antennas} (Min-UA) \textcolor{black}{$N_{T}=2N$}
is first considered with a variable-gain AF relay. It is noted that
after user scheduling the Min-UA transmission adopts a joint beamforming
to achieve SSA at the relay, where the three selected users and the
relay need to {\color{black}know} the three-party CSI. Aiming at
reducing CSI overhead, the user antenna is increased as $N_{T}=3N$,
and the transmission with an \emph{equal} number of \emph{relay} and
\emph{user} \emph{antenna} (ER-UA) is proposed with a fixed-gain AF
relay. The ER-UA transmission allows distributed beamforming at the
user with local CSI, which reduces the CSI overhead. In contrast to
the centralized scheduling schemes \cite{Taesang,upadhyay2011outage,Jeon2012opp,Joung2012user,Wang2012,Jang2010P,Xiang2012,GaoICC},
the proposed schemes can distribute the computations of scheduling
metrics to the users with local CSI. Therefore, the scheduling center
enjoys very low implementation complexity without global CSI.

The objective of this paper is to study low-complexity distributed
CS and GS for MIMO-Y channel with both Min-UA and ER-UA transmissions.
Specifically, the key contributions are summarized as follows.

1) A novel reference signal space (RSS) is proposed to guide the distributed
scheduling with both Min-UA and ER-UA transmissions. The RSS is a
predefined signal space which is known to all the nodes in the network.
Under the guidance of RSS, each user can calculate its individual
scheduling metric with local CSI, which enables several distributed
scheduling schemes with global benefits.

2) RSS-based distributed CS and GS are proposed for Min-UA transmission.
Noting that the optimal CS and GS are not decomposable for distributed
implementations with Min-UA transmission \cite{GaoICC}, two sub-optimal
angle-based scheduling strategies are proposed with RSS, which enable
distributed implementations of CS and GS. Specifically, each user
can calculate its angle/chordal-distance coordinate within the RSS by using only local CSI, and the coordinate is used to
infer the relative positions of the pair-wisely aligned signal vectors/spaces within the SSA-resultant signal space at the relay.
It is interesting to note that the selected users can generate a near-orthogonal
SSA-resultant signal space at the relay when $N=1$ and can better shape the SSA-resultant signal space when $N>1$, which results in improved
system performance.

3) RSS-based distributed CS and GS are proposed for ER-UA transmission.
Aiming at utilizing only local CSI, RSS is used to guide both distributed
beamforming and scheduling with ER-UA transmission. In particular,
each user can calculate its beamforming matrix as well as the individual
scheduling metric with local CSI and RSS. It is noted that the locally
calculated individual scheduling metric is equivalent to the link
gain of MIMO-Y channel. Therefore, such individual metric has a straightforward
connection to the optimal (centralized) scheduling metric that is
a function of all the link gains. By using the local and individual
scheduling metric, effective distributed scheduling schemes are shown
to achieve near-optimal performances.

4) The performances of the proposed schemes are analyzed. Specifically,
RSS-based distributed CS and GS are carefully analyzed for ER-UA transmission when $ N_T=N_R=3$, because of their near-optimal performances and tractability. It is
interesting to note that the distributed scheduling achieves the same
MuD order as the centralized scheduling under ER-UA transmission.
This observation is theoretically proved by studying the network's
outage probabilities and the achievable diversity-multiplexing tradeoffs
(DMTs) \cite{zheng2003diversity} with both centralized and distributed
scheduling schemes. The explicit MuD orders are obtained as $d_{CS}^{*}=\min(M_{1},M_{2},M_{3})$
for both distributed and centralized CS, and $d_{GS}^{*}=M$ for both
distributed and centralized GS, where $M_{k}$ is the number of candidates
in the $k$-th cluster $k\in\left\{ {1,2,3}\right\} $, and $M$ is
the number of candidate groups. {\color{black}Considering the former
works in \cite{GaoICC,HuiGCworkshop}, these results not only prove
the optimality of the proposed distributed scheduling in terms of
MuD order, but also shed light into the MuD behaviors in MIMO-Y channel.}

\emph{Organization}: Section II introduces the system model and the general MIMO-Y transmission. Section III describes the distributed CS and GS with Min-UA transmission. Section III details the ER-UA transmission and the corresponding distributed CS and GS, and Section IV analyzes the outage probabilities and the achievable DMTs. Numerical results and brief complexity analysis are summarized in Section VI, and Section VII concludes this paper.

{\emph{Notations}: The integer set $\left\{ 1,2,\ldots,K\right\} $
is abbreviated as $\left[1,\, K\right]$. $[\mathbf{A}]_{m,n}$, ${\left(\mathbf{A}\right)^{*}}$,
${\left(\mathbf{A}\right)^{T}}$, ${\left(\mathbf{A}\right)^{H}}$,
${\left(\mathbf{A}\right)^{-1}}$, $\mathrm{Tr\left(\mathbf{A}\right)}$,
$\mathrm{vec}\left(\mathbf{A}\right)$, $\mathrm{Range}(\mathbf{A})$, $\lambda_{min}\left(\mathbf{A}\right)$ and $\left\Vert \mathbf{A}\right\Vert _{F}$ are the $(m,n)$-th entry,
conjugate, transpose, conjugate transpose, inverse, trace, vectorization,
range, minimum eigenvalue and $F$-norm of a matrix $\mathbf{A}$. $\mathbf{I}_{m\times m}$
and $\mathbf{0}_{m\times m}$ represent the $m\times m$ identity
matrix and all-zero matrix. ${\mathop{\rm Span}\nolimits}\left({\bf {a}},{\bf {b}}\right)$
denotes the subspace spanned by vectors ${\bf {a}}$ and ${\bf {b}}$,
${\left\Vert \mathbf{a}\right\Vert }$ and $\left\langle \mathbf{a}\right\rangle =\mathbf{a}/\left\Vert \mathbf{a}\right\Vert $
are the Euclidean-norm and the normalization operation of vector $\mathbf{a}$.
$\angle\left(\mathbf{a},\mathbf{b}\right)=\cos^{-1}\left(\frac{\left|\mathbf{a}^{H}\mathbf{b}\right|}{\left\Vert \mathbf{a}\right\Vert \left\Vert \mathbf{b}\right\Vert }\right)$
is the acute angle between vector $\mathbf{a}$ and $\mathbf{b}$.
$\mathbb{C}$ represents the set of complex numbers. $\mathcal{CN}\left({\mathbf{\mathbf{m}},\mathbf{\mathbf{\Sigma}}}\right)$
denotes a complex Gaussian random vector with mean $\mathbf{\mathbf{m}}$
and covariance matrix $\mathbf{\mathbf{\Sigma}}$. $\mathrm{E}\left\{ {\cdot}\right\} $
is the expectation operator. $\tbinom{n}{k}$ is the number of $k$-combinations
from a given set of $n$ elements. $\dot{=}$ is the exponential equality,
e.g., $f\left(x\right)\dot{=}{x^{a}}$ represents $a=\mathop{\lim}\limits _{x\to\infty}\frac{{\log\left({f\left(x\right)}\right)}}{{\log\left(x\right)}}$.}

\section{System Model, MIMO-Y Transmission and RSS}

\subsection{System Model and MIMO-Y Transmission}
As shown in Fig. 1, a MIMO-Y network comprises an $N_{R}$-antenna
relay $\mathsf{R}$ and three clusters of {\color{black}$N_{T}$-antenna
users $\left\{ \mathsf{S}_{j_{k}},\,{j}\in\left[{1,{M_{k}}}\right],\, k\in[1,3]\right\} $},
$j_{k}$ and $M_{k}$ are the intra-cluster user index and the number
of candidates within the $k$-th cluster. It is assumed that there is no direct link between any two users
in different clusters, and the half-duplex AF relay helps information
exchange among clusters. Time-division duplex (TDD) mode is assumed,
therefore channel reciprocity holds. {\color{black}The channels of $\mathsf{\mathsf{S}}_{j_{k}}\rightarrow\mathsf{R}$
and $\mathsf{R\rightarrow\mathsf{S}}_{j_{k}}$ are denoted as {\color{black}$\mathbf{H}_{j_{k}}\in\mathbb{C}^{N_{R}\times N_{T}}$
and $\mathbf{H}_{j_{k}}^{H}\in\mathbb{C}^{N_{T}\times N_{R}}$}},
respectively\footnote{Rigourously speaking, the channel of $\mathsf{R\rightarrow\mathsf{S}}_{j_{k}}$ should be $\mathbf{H}_{j_{k}}^{T}$ instead of $\mathbf{H}_{j_{k}}^{H}$. However, according to the conjugate operations in \cite{Zhiguo_MP}, the two channel models can be equivalent for performance analysis.}, whose entries are independent and identically distributed
(i.i.d.) $\mathcal{CN}\left(0,\,1\right)$. User scheduling is the
focus of this paper, and only one user is selected from each cluster.
The three selected users then exchange information through the basic
MIMO-Y channel {\cite{lee2010degrees}}.{ \color{black} More specifically, each user sends two unicast messages for the other two users, and intends to decode two messages from the other users.} Before presenting the specific
transmission and scheduling schemes, the outline of the general MIMO-Y
transmission is reviewed in this section. For the ease of exposition,
the intra-cluster index of each user is temporarily neglected, and
the selected user in the $k$-th cluster is denoted as $\mathsf{S}_{k}$.
Accordingly, the relevant channels of $\mathsf{S}_{k}$ are denoted
as $\mathbf{H}_{k}$ and $\mathbf{H}_{k}^{H}$, respectively. In addition,
the information symbols from $\mathsf{S}_{k}$ to another two users
$\left\{ \mathsf{S}_{l}\right\} _{l\text{\ensuremath{\in}}\mathcal{L}_{k}}$
are collected in $\left\{ \mathbf{d}_{l,k}\right\} _{l\text{\ensuremath{\in}}\mathcal{L}_{k}}$,
$\mathcal{L}_{k}=\left[1,3\right]\backslash\left\{ k\right\} $, {\color{black}where
$\mathbf{d}_{l,k}=\left[d_{l,k}^{\left[1\right]},d_{l,k}^{\left[2\right]},\ldots,d_{l,k}^{\left[N\right]}\right]^{T}\in\mathbb{C}^{N\times1}$
is the unicast of $\mathsf{S}_{k}\rightarrow\mathsf{S}_{l}$ containing
$N$ data streams.}
\begin{figure}[!t]
\centering \includegraphics[width=9cm]{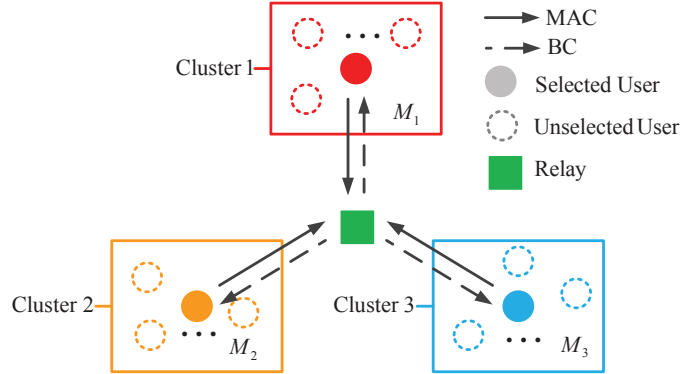}\label{FIG 1}
\protect\caption{User scheduling for MIMO-Y channel, where one user is selected from
each cluster to exchange information via a relay. The ANC-based protocol
is used which consists of MAC and BC phases.}
\end{figure}
Analog network coding (ANC) \cite{SigANC} is employed for efficient
AF relaying, which consists of the multiple access (MAC) and broadcasting
(BC) transmission phases. {\color{black}In the MAC phase, ANC treats the superimposed signals as network-coded symbols and just amplify-and-forwards them in the BC phase. Upon receiving the broadcasted symbols from the relay, users extract the desired signals by virtue of self-interference cancelation.} In order to exploit ANC for MIMO-Y transmission, during the MAC phase, user $\mathsf{S}_{k}$
uses the transmit beamforming matrix {\color{black} $\mathbf{V}_{k}=[\mathbf{V}_{l_{1},k}\,\mathbf{V}_{l_{2},k}]\in\mathbb{C}^{N_{T}\times2N}$}
for data, and the transmitted symbol vector is $\mathbf{s}_{k}=\mathbf{V}_{k}\mathbf{d}_{k}\in\mathbb{C}^{N_{T}\times1}$,
where $\mathbf{d}_{k}=[\mathbf{d}_{l_{1},k}^{T}\,\mathbf{d}_{l_{2},k}^{T}]^{T}\in\mathbb{C}^{2N\times1}$
and $\mathrm{E}\left\{ \mathbf{d}_{k}\mathbf{d}_{k}^{H}\right\} =\mathbf{I}_{2N\times2N}$,
$k\in\left[1,3\right]$, $l_{1},\, l_{2}\in\mathcal{L}_{k}$, $l_{1}\neq l_{2}$.
A transmit power constraint is imposed as $\mathrm{Tr}\left(\mathbf{V}_{k}\mathbf{V}_{k}^{H}\right)\leq P_{T}$,
where $P_{T}$ is the average transmit power of each user. Then, the
relay $\mathsf{R}$ receives
{\small\begin{equation}
\mathbf{y}_{R}=\sum_{k=1}^{3}\mathbf{H}_{k}\mathbf{s}_{k}+\mathbf{n}_{R},\label{eq:M_Rx_Sig}
\end{equation}}where $\mathbf{n}_{R}\in\mathbb{C}^{N_{R}\times1}\sim\mathcal{CN}\left(\mathbf{0},\,\sigma_{R}^{2}\mathbf{I}_{N_{R}\times N_{R}}\right)$ is the additive
white Gaussian noise (AWGN) vector.
During the BC phase, the transmitted signal of the AF relay is given
as $\mathbf{s}_{R}=\mathbf{W}\mathbf{y}_{R}=G_{R}\mathbf{\tilde{W}}\mathbf{y}_{R}\in\mathbb{C}^{N_{T}\times1}$,
where $\mathbf{\tilde{W}}\in\mathbb{C}^{N_{R}\times N_{R}}$ is the
relay processing matrix and $G_{R}$ is the power controlling coefficient
to be specified later. {\color{black} Then, the received
signal at $\mathsf{S}_{k}$ is expressed as
{\small\begin{equation}
\mathbf{y}_{k}=\mathbf{H}_{k}^{H}\mathbf{s}_{R}+\mathbf{n}_{k}\label{eq:M_Sx_Sig}
\end{equation}}where $\mathbf{n}_{k}\in\mathbb{C}^{N_{T}\times1}$
is the AWGN vector distributed as $\mathcal{CN}\left(\mathbf{0},\,\sigma_{S}^{2}\mathbf{I}_{N_{T}\times N_{T}}\right)$.}
According to the ANC protocol, $\mathsf{S}_{k}$ needs to perform
self-interference cancellation before extracting the useful information
sent by $\left\{ \mathsf{S}_{l}\right\} _{l\text{\ensuremath{\in}}\mathcal{L}_{k}}$
with the receive beamforming matrix $\mathbf{U}_{k}\in\mathbb{C}^{N_{T}\times2N}$,
which is described as
{\small\begin{align}
\mathbf{\hat{y}}_{k} & =\mathbf{U}_{k}^{H}\left(\mathbf{y}_{k}-\mathbf{H}_{k}^{H}\mathbf{W}\mathbf{H}_{k}\mathbf{s}_{k}\right)=\mathbf{\mathbf{\hat{y}}}_{k,S}+\mathbf{\mathbf{\hat{y}}}_{k,\sigma},\label{eq:D_Self_IC}
\end{align}}{\color{black}where $\mathbf{\mathbf{\hat{y}}}_{k,S}=\mathbf{U}_{k}^{H}\mathbf{H}_{k}^{H}\mathbf{W}\sum_{l\in\mathcal{L}_{k}}\mathbf{H}_{l}\mathbf{V}_{l}\mathbf{d}_{l}\in\mathbb{C}^{2N\times1}$
 and $\mathbf{\mathbf{\hat{y}}}_{k,\sigma}=\mathbf{U}_{k}^{H}\left(\mathbf{H}_{k}^{H}\mathbf{W}\mathbf{n}_{R}+\mathbf{n}_{k}\right)\in\mathbb{C}^{2N\times1}$}
are the signal component and the noise component of the decision variable $\mathbf{\hat{y}}_{k}$, respectively.
With a proper design of transmit/receive beamforming matrices $\left\{ \mathbf{U}_{k},\,\mathbf{V}_{k}\right\} _{k=1}^{3}$
at {\color{black}each user} and $\mathbf{W}$ at the relay, $\mathsf{S}_{k}$
can have interference-free reception and recover the useful information
as $\left(\hat{\mathbf{d}}_{k,l_{1}}\,\hat{\mathbf{d}}_{k,l_{2}}\right)=f_{k}\left(\mathbf{\hat{y}}_{k}\right),\, l_{1},\, l_{2}\in\mathcal{L}_{k}$,
$l_{1}\neq l_{2}$, where {\color{black}$f_{k}$ represents the decoding
process at $\mathsf{S}_{k}$}. In the next sections, we will show
the detailed designs of $\left\{ \mathbf{U}_{k},\,\mathbf{V}_{k}\right\} _{k=1}^{3}$
and $\mathbf{W}$ for Min-UA and ER-UA transmissions as well as their
corresponding CS and GS. {\color{black}It is also noted that we mainly
focus on two typical antenna configurations, namely $\frac{N_{R}}{N_{T}}=\frac{3}{2}$
and $\frac{N_{R}}{N_{T}}=1$ with $N_{R}=3N$, which represents the
Min-UA transmission and the ER-UA transmission, respectively. Here, $N\geq1$ is the number of data stream(s) of each unicast transmission within the MIMO-Y channel, and the minimum number of relay antennas to support such MIMO-Y transmission is exactly $3N$.}

\subsection{RSS}

{\color{black} One of the key contribution of this paper is a novel RSS introduced for user scheduling. To be more specific,
the RSS is a predefined signal space $\Omega_{R}=\mathrm{Span}\left(\left\{ \mathbf{e}_{{\rm I}}^{[n]},\mathbf{e}_{{\rm II}}^{[n]},\mathbf{e}_{{\rm III}}^{[n]}\right\} _{n=1}^{N}\right)$,
whose normalized orthogonal basis $\mathbf{E}=\left[\mathbf{E}_{{\rm I}}\,\mathbf{E}_{{\rm II}}\,\mathbf{E}_{{\rm III}}\right]\in\mathbb{C}^{N_{R}\times N_{R}}$
is assumed to be known by all the users in this network, where $\mathbf{E}_{m}=\left[\mathbf{e}_{m}^{[1]},\mathbf{e}_{m}^{[2]},\ldots,\mathbf{e}_{m}^{[N]}\right]\in\mathbb{C}^{N_{R}\times N}$
can span a subspace%
\footnote{For notational clarity, a permutation function over the source index
pair $(l,\, k)$ is introduced as $m=\pi{(l,\, k)}=\pi{(k,\, l)}$,
$m,\, k\in\left[1,3\right]$, $l\in\mathcal{L}_{k}$. Specifically,
$\pi$ is defined as $\pi\left(1,\,2\right)={\rm I}$, $\pi\left(1,\,3\right)={\rm II}$,
$\pi\left(2,\,3\right)={\rm III}$. %
}, $m\in\left\{ \mathrm{I},\mathrm{II},\mathrm{III}\right\}$. It is noted that, $\mathbf{E}$ can be arbitrary normalized orthogonal
basis of the $N_{R}$-dimensional space and can be designed off-line
or broadcasted by the relay. The RSS $\mathbf{E}$ is used to guide distributed
user scheduling with Min-UA and ER-UA transmissions. In particular,
for the Min-UA transmission, the RRS $\mathbf{E}$ will be used to
shape the signal space seen by the relay or the SSA-resultant signal
space $\mathbf{F}$. It will be shown later, because less antennas
are equipped at the user, the user is not able to perfectly align
its signal space with the predefined direction or subspace of $\mathbf{E}$.
Therefore, the SSA-resultant signal space $\mathbf{F}$
can only be shaped by RSS-based scheduling for the Min-UA transmission, that is to say $\mathbf{F}$ can not be totaly determined by $\mathbf{E}$. On the other hand, for the ER-UA transmission,
each user is equipped with enough antennas to perfectly
align its transmit signal space with the predefined direction or subspace of the RSS $\mathbf{E}$.
Therefore, the SSA-result signal space $\mathbf{F}$ at relay can be exactly determined
by $\mathbf{E}$ with ER-UA transmission.}

\section{Distributed User Scheduling with Min-UA Transmission}

In this section, the distributed user scheduling schemes are studied
with the Min-UA transmission, where each user is equipped with $N_{T}=2N$
antennas and the relay is equipped with $N_{R}=3N$ antennas, $N\geq1$.
It is noted that in this scenario the instantaneous three-party CSI
is required by the three selected users and the relay for joint beamforming
\cite{lee2010degrees}. Because of this coupling, the calculation
of the optimal scheduling metric, i.e., the post-processing signal-to-noise-ratio
(SNR), cannot be easily decomposed, and the design of an effective
distributed user scheduling is very challenging. To this end, the
RSS $\mathbf{E}$ is introduced to guide the distributed CS and GS. Before presenting the RSS-based distributed
user scheduling schemes, the Min-UA transmission is briefly {\color{black}described}
in the following subsection.

\subsection{Min-UA Transmission}

Without loss of generality, let us assume three users $\left\{ \mathsf{S}_{1},\,\mathsf{S}_{2},\,\mathsf{S}_{3}\right\} $
are randomly selected to perform Min-UA MIMO-Y transmission. \textcolor{black}{In
the MAC phase, each user sends 2 unicast messages which consists of
$2N$ independent data streams. Therefore, there are $6N$ data streams
arriving at the relay simultaneously. Since the relay is only equipped
with $N_{R}=3N$ antennas, it is not able to decode these signals
in a stream-by-stream fashion. In order to utilize the relay antennas
more efficiently, SSA is introduced into the MIMO Y channel.} In particular,
SSA is required for the bi-directional information exchange between
the pair $\mathsf{S}_{l}$ and $\mathsf{S}_{k}$, $k\in[1,3]$, $l\in\mathcal{L}_{k}$,
and the transmit beamforming matrix of each user is jointly designed
with another two users using the three-party CSI. More specifically,
the pair-wise transmit beamforming matrices of $\mathsf{S}_{l}$ and
$\mathsf{S}_{k}$ can be \emph{jointly} designed by solving the null-space
problem \cite{lee2010degrees} as
{\small\begin{equation}
\left[\mathbf{H}_{l}\,\,\,-\mathbf{H}_{k}\right]\mathbf{\tilde{V}}_{m=\pi\left(l,k\right)}=\mathbf{0},\label{eq:SSA-EQ}
\end{equation}}where \textcolor{black}{$\mathbf{\tilde{V}}_{m}=\left[\begin{array}{cc}
\mathbf{\tilde{V}}_{k,l}^{T} & \mathbf{\tilde{V}}_{l,k}^{T}\end{array}\right]^{T}\in\mathbb{C}^{{\color{black}2N_{T}\times N}}$}
contains the pair-wise transmit beamforming matrices, $\mathbf{\tilde{V}}_{k,l}$ and $\mathbf{\tilde{V}}_{l,k}$, with a total
power normalization as ${\color{black}\left\Vert \mathbf{\tilde{V}}_{m}\right\Vert ^{2}_{F}=1}$.
For simplicity, a total power constraint $P_{T}$ is imposed on this
pair-wise transmit beamforming matrices, and the effective transmit
beamforming matrix for \textcolor{black}{$\mathbf{d}_{l,k}$} is expressed
as
{\small\begin{equation}
{\color{black}\mathbf{V}_{l,k}=\sqrt{P_{T}}\mathbf{\tilde{V}}_{l,k}.}\label{eq:TB-SSA}
\end{equation}}It is easy to check that each unicast message of $\mathsf{S}_{k}$
has an average power of \textcolor{black}{$\mathrm{E}\left\{ \left\Vert \mathbf{V}_{l,k}\right\Vert ^{2}_{F}\right\} =\frac{1}{2}P_{T}$},
therefore, the average transmit power of each user is $P_{T}$. According
to (\ref{eq:SSA-EQ}) and (\ref{eq:TB-SSA}), it is noted that the
three-party CSI is necessary for the beamforming matrix \textcolor{black}{$\mathbf{V}_{k}=[\mathbf{V}_{l_{1},k}\,\mathbf{V}_{l_{2},k}]$}
of $\mathsf{S}_{k}$, $l_{1},\, l_{2}\in\mathcal{L}_{k}$, $l_{1}\neq l_{2}$.
Employing the pair-wise transmit beamforming matrix \textcolor{black}{$\mathbf{V}_{k,l}$}
and \textcolor{black}{$\mathbf{V}_{l,k}$}, the bi-directional signal
between $\mathsf{S}_{l}$ and $\mathsf{S}_{k}$ is then aligned within
a $N$-dimensional subspace spanned by the column vectors of \textcolor{black}{$\mathbf{F}_{m=\pi\left(k,l\right)}=\mathbf{H}_{l}\tilde{\mathbf{V}}_{k,l}=\mathbf{H}_{k}\mathbf{\tilde{V}}_{l,k}\in\mathbb{C}^{N_{R}\times N}$},
which is within the signal space of $\mathsf{R}$, and the received
signal at $\mathsf{R}$ is given by (cf. (\ref{eq:M_Rx_Sig}))
{\small\begin{equation}
\mathbf{y}_{R}=\sqrt{P_{T}}\mathbf{\mathbf{F}}\mathbf{d}_{+}+\mathbf{n}_{R},\label{eq:SSA-R}
\end{equation}}where {\textcolor{black}{$\mathbf{F}=\left[\mathbf{F}_{{\rm I}}\,\mathbf{F}_{{\rm II}}\,\mathbf{F}_{{\rm III}}\right]\in\mathbb{C}^{N_{R}\times N_{R}}$}
is the signal space seen by $\mathsf{R}$ with $\mathbf{F}_{m}=\left[\mathbf{f}_{m}^{[1]},\mathbf{f}_{m}^{[2]},\ldots,\mathbf{f}_{m}^{[N]}\right]\in\mathbb{C}^{N_{R}\times{N}}$,}and \textcolor{black}{$\mathbf{d}_{+}=\left[\mathbf{d}_{+,{\rm I}}^{T}\,\mathbf{d}_{+,{\rm II}}^{T}\,\mathbf{d}_{+,{\rm III}}^{T}\right]^{T}\in\mathbb{C}^{3N\times1}$}
is the superimposed signal with element \textcolor{black}{$\mathbf{d}_{+,m}=\mathbf{d}_{l,k}+\mathbf{d}_{k,l}$}. Taking $N_{R}=3$ as an example%
\footnote{For the ease of illustration, most of the figures are based on the
assumption that $N_{R}=3$, accordingly, $\mathbf{E}$ is simplified
as $\mathbf{E}=[\mathbf{e}_{\mathrm{I}},\mathbf{e}_{\mathrm{II}},\mathbf{e}_{\mathrm{III}}]\in\mathbb{C}^{3\times3}$
and $\mathbf{F}$ is simplified as $\mathbf{F}=[\mathbf{f}_{\mathrm{I}},\mathbf{f}_{\mathrm{II}},\mathbf{f}_{\mathrm{III}}]\in\mathbb{C}^{3\times3}$
for these figures. %
}, Fig. 2 describes the idea of SSA.
For simplicity, we follow \cite{ding2011generalized} and use the
zero forcing (ZF)-based relay processing matrix \textcolor{black}{$\mathbf{W}=G_{R}\mathbf{\tilde{W}}=G_{R}\left(\mathbf{\mathbf{\mathbf{F}}}^{H}\right)^{-1}\mathbf{{\mathbf{F}}}^{-1}=G_{R}\left(\mathbf{F}\mathbf{F}^{H}\right)^{-1}\in\mathbb{C}^{N_{R}\times N_{R}}$
}at $\mathsf{R}$, \textcolor{black}{where $\mathbf{{\mathbf{F}}}^{-1}$
is the detection matrix and $\left(\mathbf{\mathbf{\mathbf{F}}}^{H}\right)^{-1}$
is the transmit beamforming matrix. It is noted that $\tilde{\mathbf{W}}$
decouples the equivalent channel matrices of the MAC and BC phases
respectively, so that the users can obtain the desired superimposed signals
or network-coded signals at low cost. Here, $G_{R}$ is the power
controlling coefficient of the variable-gain AF relay, and it is calculated
as
{\small\begin{equation}
G_{R}=\sqrt{P_{R}/\mathsf{E}\left\{ \left\Vert \mathbf{x}_{R}\right\Vert ^{2}\right\} }=\sqrt{P_{R}/\left(P_{T}\mathrm{Tr}\left(\mathbf{\tilde{W}}\right)+\sigma_{R}^{2}\mathrm{Tr}\left(\mbox{\ensuremath{\mathbf{\tilde{W}}}}\mathbf{\tilde{W}}^{H}\right)\right)},\label{eq:gain-RSS}
\end{equation}}where $\mathbf{x}_{R}=\mathbf{W}\mathbf{y}_{R}=G_{R}\sqrt{P_{T}}\left(\mathbf{F}^{H}\right)^{-1}\mathbf{d}_{+}+G_{R}\tilde{\mathbf{W}}\mathbf{n}_{R}$
is the signal broadcasted by the relay, and the expectation on $\left\Vert \mathbf{x}_{R}\right\Vert ^{2}$
is over $\mathbf{d}_{+}$ and $\mathbf{n}_{R}$. }

\begin{figure}[t]
\centering \includegraphics[width=9cm]{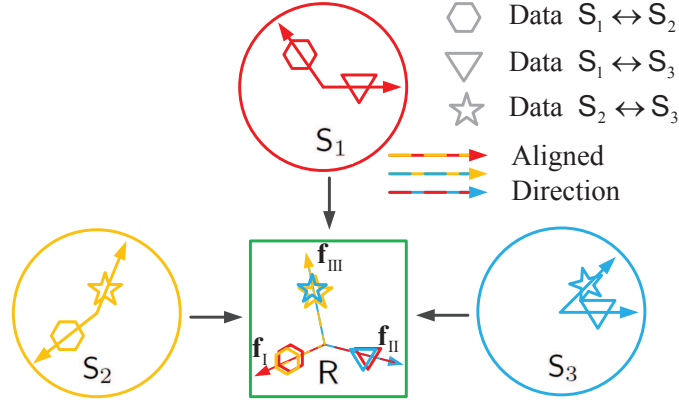} \protect\caption{Geometrical interpretation on the Min-UA transmission. After the joint
beamforming, the signals are pair-wisely aligned at the relay during
the MAC phase. The aligned directions are not necessarily orthogonal,
and the equivalent MIMO channel seen by the relay might be ill-conditioned.}
\end{figure}

At each user $\mathsf{S}_{k}$, the received signal is given by (\ref{eq:M_Sx_Sig}),
then the remaining signal after subtracting self-interference is given
by
{\small\begin{align}
\mathbf{\tilde{y}}_{k} & =\sqrt{P_{T}}\mathbf{H}_{k}^{H}\mathbf{W}\mathbf{{\mathbf{F}}}_{k}\mathbf{d}_{+,k}+\mathbf{H}_{k}^{H}\mathbf{W}\mathbf{n}_{R}+\mathbf{n}_{k},\label{eq:SSA substract self}
\end{align}}where ${\color{black}\mathbf{{\mathbf{\tilde{F}}}}_{k}=\left[\mathbf{{F}}_{m_{1}}\,\mathbf{{F}}_{m_{2}}\,\mathbf{{F}}_{m'}\right]\in\mathbb{C}^{N_{R}\times N_{R}}}$,
and \textcolor{black}{$\mathbf{\mathbf{d}}_{+,k}=\left[\mathbf{d}_{k,l_{1}}^{T}\,\mathbf{d}_{k,l_{2}}^{T}\,\mathbf{d}_{+,m'}^{T}\right]^{T}\in\mathbb{C}^{3N\times1}$},
$m_{1}=\pi\left(l_{1},k\right)$, $m_{2}=\pi\left(l_{2},k\right)$,
$l_{1},\, l_{2}\in\mathcal{L}_{k}$ and $m'\in\left[1,3\right]\backslash\left\{ m_{1},m_{2}\right\} .$
Then the receive beamforming \textcolor{black}{$\mathbf{U}_{k}=\left[\mathbf{U}_{l_{1},k}\,\mathbf{U}_{l_{2},k}\right]=\left[\mathbf{V}_{l_{1},k}\,\mathbf{V}_{l_{2},k}\right]\in\mathbb{C}^{N_{T}\times2N}$}
is applied to obtain $\mathbf{\hat{y}}_{k}=\mathbf{U}_{k}^{H}\tilde{\mathbf{y}}_{k}$.
\textcolor{black}{More specifically, $\mathbf{U}_{l,k}=\mathbf{V}_{l,k}$
is used to get the signal from $\mathsf{S}_{l}$ as $\hat{\mathbf{y}}_{k,l}=\mathbf{U}_{l,k}^{H}\tilde{\mathbf{y}}_{k}$},
and each desired signal stream within \textcolor{black}{$\hat{\mathbf{y}}_{k,l}$
}is given by
{\small\begin{equation}
{\color{black}{\hat{y}}_{k,l}^{[n]}=\sqrt{P_{T}}G_{R}d_{k,l}^{[n]}+G_{R}\mathbf{a}_{n}^{T}{\mathbf{F}_{m}^{H}}\mathbf{\tilde{W}}\mathbf{n}_{R}+\mathbf{a}_{n}^{T}\mathbf{V}_{l,k}^{H}\mathbf{n}_{k},\, n\in[1,N],}\label{eq:EQ-P2P-SSA}
\end{equation}}\textcolor{black}{where the elements of $\mathbf{a}_{n}\in\mathbb{C}^{N\times1}$
are all zeros except the $n$-th element that equals to one, i.e.,
$\mathbf{a}_{n}$ is a Cartesian unit vector.} After briefing the Min-UA
transmission, the corresponding user scheduling schemes are discussed
in the next subsection.

\subsection{Problem Formulation and Centralized Scheduling}

In this subsection, the optimal or centralized CS and GS are formulated with Min-UA
transmission, which serves as a preliminary and reference for distributed
scheduling. Some necessary notations and performance metrics are first
introduced. For ease of description on CS, an ordered set $J=\left\{ j_{1},j_{2},j_{3}\right\} $
is used as a collection of the user indices%
\footnote{It is noted that the
order in $J$ can indicate the cluster $k$. Therefore, we can drop the subscript $k$ of $j_{k}$ to simplify the notation of the intra-cluster index within $J$, if not causing confusion. For example, instead of using $J=\{3_1,2_2,1_3\}$, we can use $J=\{3,2,1\}$ to
represent the 3rd, the 2nd and the 1st users from the 1st, 2nd, 3rd
clusters, respectively.%
}, $j\in[1,\, M_{k}],$ $k\in[1,3]$, and all the possible realizations
of $J$ are enumerated%
\footnote{We can order by the first, then by the second, and then by the third
elements of $J$. For example, $J_{1}=\left\{ 1,1,1\right\} \prec J_{2}=\left\{ 2,1,1\right\} \prec\ldots\prec J_{q}=\left\{ j_{1},j_{2},j_{3}\right\} \prec\ldots\prec J_{M_{1}M_{2}M_{3}}=\left\{ M_{1},M_{2},M_{3}\right\} $.%
} in another ordered set $\mathcal{J}=\{J_{q}\}_{q=1}^{Q}$, $Q=M_{1}M_{2}M_{3}$.
\textcolor{black}{In GS, we have $M_{1}=M_{2}=M_{3}=M$. Then, a subset $\mathcal{J}'\subset\mathcal{J}$
is used to facilitate the description for all the possible realizations
of $J'_{p}=\left\{ j'_{1},j'_{2},j'_{3}|j'_{1}=j'_{2}=j'_{3}=p\right\} $,
which is defined as%
\footnote{It is noted that the index used for GS could have been subsumed under
$J$ or simplified with less symbols, however, we keep this seemingly
redundant form to unify the descriptions on user scheduling criteria
and the subsequential analysis.%
} $\mathcal{J}'=\{J'_{p}\}_{p=1}^{M}$}. Regarding the system performance
metrics, the overall outage probability of the network with arbitrary
selected users $\left\{ \mathsf{S}_{j_{1}},\,\mathsf{S}_{j_{2}},\,\mathsf{S}_{j_{3}}\right\} $
is defined as
{\small\begin{equation}
P_{out}^{\left(J\right)}\left(\rho_{th}\right)=\Pr\left(\rho_{min}^{\left(J\right)}\leq\rho_{th}\right),\label{eq:Pout}
\end{equation}}where {\color{black}$\rho_{th}$ is the threshold SNR-value for the
outage probability and} $\rho_{min}^{\left(J\right)}$ is the minimum
post-processing SNR (min-SNR) of the network with selected users in
$J$. Then, $\rho_{min}^{\left(J\right)}$ is defined as
{\small\begin{equation}
\rho_{min}^{\left(J\right)}=\min_{k\in[1,3],\, l\text{\ensuremath{\in}}\mathcal{L}_{k},\thinspace n\in[1,N]}\left(\rho_{j_{k},j_{l}}^{[n]}\right),\label{eq:OR_Pout}
\end{equation}}{\color{black}where $\rho_{j_{k},j_{l}}^{[n]}$ represents the end-to-end
post-processing SNR of the $n$-th data stream of link $\mathsf{\mathsf{S}}_{j_{l}}\rightarrow\mathsf{R}\rightarrow\mathsf{\mathsf{S}}_{j_{k}}$,
$l\text{\ensuremath{\in}}\mathcal{L}_{k}$, and $\rho_{j_{k},j_{l}}^{[n]}=\mathrm{\mathsf{E}}\left\{ \left(\sqrt{P_{T}}G_{R}d_{k,l}^{[n]}\right)^{2}\right\} /\mathsf{E}\left\{ \left(G_{R}\mathbf{a}_{n}^{T}{\mathbf{F}_{m}^{H}}\mathbf{\tilde{W}}\mathbf{n}_{R}+\mathbf{a}_{n}^{T}\mathbf{V}_{l,k}^{H}\mathbf{n}_{k}\right)^{2}\right\} $
is calculated with reference to (\ref{eq:EQ-P2P-SSA}) as
{\small\begin{equation}
\rho_{j_{k},j_{l}}^{[n]}=\frac{G_{R}^{2}P_{T}}{G_{R}^{2}\sigma_{R}^{2}\left[\mathbf{F}_{m}^{H}\mathbf{\tilde{W}}^{\left(J\right)}\left(\mathbf{\tilde{W}}^{\left(J\right)}\right)^{H}\mathbf{F}_{m}\right]_{n,n}+\sigma_{S}^{2}\left[\mathbf{V}_{j_{l},j_{k}}^{H}\mathbf{V}_{j_{l},j_{k}}\right]_{n,n}},\label{eq:new_post_processing_SNR}
\end{equation}}}where $G_{R}$ is given by (\ref{eq:gain-RSS}). Based on the aforementioned
notations, the centralized CS is given by
{\small\begin{equation}
J_{C}=\left\{ j_{1}^{*},j_{2}^{*},j_{3}^{*}\right\} =\arg\,\max_{J\in\mathcal{J}}\,\left(\rho_{min}^{\left(J\right)}\right),\label{eq:C_CS}
\end{equation}}and the centralized GS is given by
{\small\begin{equation}
J'_{C}=\left\{ j_{1}^{'*},j_{2}^{'*},j_{3}^{'*}\right\} =\arg\,\max_{J'\in\mathcal{J}'}\,\left(\rho_{min}^{\left(J'\right)}\right).\label{eq:Min-UA-C}
\end{equation}}Similar to the scheduling in \cite{GaoICC}, both centralized CS and
GS involve global CSI and high computational complexity at the scheduling
center.

\subsection{RSS-based Distributed scheduling with Min-UA for $N=1$}

\begin{figure}[t]
\centering \includegraphics[width=9cm]{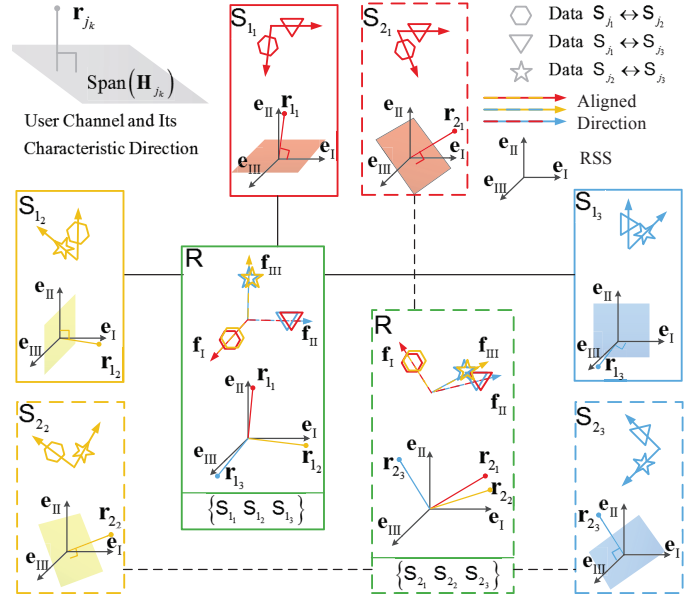} \protect\caption{The geometrical interpretations on the RSS-based distributed user
scheduling with Min-UA transmission. In this example, we consider
the cluster-wise scheduling and assume that there are two users $\left\{ {{\mathsf{S}_{{1_{k}}}},{\mathsf{S}_{{2_{k}}}}}\right\} $
in the $k$-th cluster, $k\in\left[{1,3}\right]$. The angles between
each user's {\color{black}characteristic} direction ${{\bf {r}}_{{j_{k}}}}$
and the basis of RSS $\left\{ {{{\bf {e}}_{{\rm I}}},{{\bf {e}}_{{\rm II}}},{{\bf {e}}_{{\rm III}}}}\right\} $
can be calculated by the user in a distributed manner. Employing the
proposed distributed user scheduling, the pattern of users' characteristic
directions can be directly optimized so that the SSA-resultant signal
space ${\bf {F}}$ can be indirectly shaped to be more orthogonal.}
\end{figure}

%Before presenting the specific scheduling schemes,
%the geometrical interpretations on the SSA-resultant signal space
%and RSS are reviewed to motivate our design. As shown in Fig. 3, the
%SSA-resultant signal space $\mathbf{{\mathbf{F}}}$ is composed of
%three vectors $\left\{ \mathbf{{f}}_{m}\right\} _{m={\rm I}}^{{\rm III}}$.

In this subsection, RSS-based distributed CS and GS are first proposed with
Min-UA transmission for $N=1$. {\color{black} The main
idea of the RSS-based scheduling is that each user can calculate its
scheduling metric with local CSI and feed the metric back to the relay
in a very efficient way, and then the relay can select the proper
users or user group whose SSA-resultant signal space $\mathbf{F}$
is well-shaped. More specifically, the RSS $\mathbf{E}$ is commonly
known by all the users, based on which each user can calculate the
direction (in terms of angle) of its own channel (which is in fact
a subspace) within the RSS with just local CSI. Then, the users feed
the angle-based scheduling metrics back to the relay, and the relay
can further infer the degree of orthogonality among the vectors (subspaces)
of the SSA-resultant $\mathbf{F}$ and selects the favorable users/user
group to shape $\mathbf{F}$. It is noted that if $\mathbf{F}$ can
be well-shaped, the power loss caused by the ZF-based transceiver
$\mathbf{\tilde{W}}$ can be mitigated and the system performance
can be improved.}

The interpretations on the orthogonality among channels may have impact
on the design of user scheduling. {\color{black}  A straightforward
observation is that the vectors $\left\{ \mathbf{{f}}_{m}\right\} _{m={\rm I}}^{{\rm III}}$ within the SSA-resultant signal space $\mathbf{{\mathbf{F}}}$ can be shaped to be near-orthogonal by user scheduling.} However, this observation is
not instructive for an efficient distributed scheduling with just
local CSI. It is noted that $\mathbf{{f}}_{m}$ lies in the {\color{black}intersection
space} of $\mathbf{H}_{j_{k}}$ and $\mathbf{H}_{j_{l}}$, i.e.,
$\mathrm{Span}\left(\mathbf{{f}}_{m}\right)=\mathrm{Range}\left(\mathbf{H}_{j_{k}}\right)\cap\mathrm{Range}\left(\mathbf{H}_{j_{l}}\right)$,
$m=\pi\left({l,k}\right)$, and the measurement of orthogonality
between ${{\bf {f}}_{m}}$ and ${{\bf {f}}_{m'}}$, $m'\ne m$, requires
the three-party CSI. Therefore, new methods should be developed to
enable distributed scheduling. {\color{black}Considering only local
CSI at $\mathsf{S}_{j_{k}}$}, i.e., the channel $\mathbf{H}_{j_{k}}$
of user $\mathsf{S}_{j_{k}}$, we first introduce the characteristic
direction of the channel as $\mathbf{r}_{j_{k}}\in\mathrm{Null}\left(\mathbf{H}_{j_{k}}\right)$,
$\left\Vert \mathbf{r}_{j_{k}}\right\Vert =1$. As shown in Fig. 3,
if we can make ${\left\{ {{\bf {r}}_{{j_{k}}}}\right\} _{k=1}^{3}}$
orthogonal, \textcolor{black}{then the channels ${\left\{ {{\bf {H}}_{{j_{k}}}}\right\} _{k=1}^{3}}$
are pair-wisely perpendicular in the 3-dimensional space}, and eventually the {\color{black}intersection spaces}
of theses channels, i.e., $\left\{ \mathbf{{f}}_{m}\right\} _{m={\rm I}}^{{\rm III}}$,
are orthogonal. It is interesting to note that the shaping of $\left\{ \mathbf{{f}}_{m}\right\} _{m={\rm I}}^{{\rm III}}$
can be achieved in a distributed manner with RSS and local CSI. To
make this intuitive observation more concrete, the RSS-based distributed
CS and GS are detailed in the following subsections. %It is easy to note that each user can calculate its \emph{local} angular-coordinate%within the RSS; therefore, it is possible to shape $\mathbf{{\mathbf{F}}}$%with distributed methods.

\subsubsection{RSS-based distributed CS}

Recall that the RSS is a predefined {\color{black}3-dimensional}
signal space $\Omega_{R}=\mathrm{Span}\left(\mathbf{e}_{{\rm I}},\mathbf{e}_{{\rm II}},\mathbf{e}_{{\rm III}}\right)$,
which is known by all the users in the network. Then, each user can
calculate the angular-coordinate of its characteristic direction within
RSS with only local CSI as $\phi_{j_{k}}=[\phi_{j_{k},{\rm I}}\,\phi_{j_{k},{\rm II}}\,\phi_{j_{k},{\rm III}}]$,
where $\phi_{j_{k},m}=\angle\left(\mathbf{r}_{j_{k}},\mathbf{e}_{m}\right)$.
The proposed RSS-based distributed CS aims to find the best user from
each cluster based on the angular-coordinate. Specifically, the RSS-based
distributed CS is a three-round sequential scheduling scheme. Let
$\mathcal{K}\left(m\right)\subseteq[1,3]$ be the set of candidate
cluster-{\color{black}indices} for the $m$-th round, $m\in\left\{ {\rm I},{\rm II},{\rm III}\right\} $.
Then, the procedure of distributed CS with Min-UA transmission is
summarized as follows:
\begin{enumerate}
\item Initialization: Let $\mathcal{K}\left(1\right)=[1,3]$.
\item User selection: During the $m$-th round, the user whose characteristic
direction is mostly aligned with $\mathbf{e}_{m}$ is selected from
the $\left\{ k\in\mathcal{K}\left(m\right)\right\} $ cluster(s) as
{\small\begin{equation}
j_{\mu\left(m\right)}^{\ddagger}=\arg\underset{\left\{ {{j_{k}}|{j}\in[1,{M_{k}}],\; k\in\mathcal{K}\left(m\right)}\right\} }{\min}\left(\phi_{j_{k},m}\right),\label{eq:D_CS}
\end{equation}}
where $\mu\left(m\right)$ represents the cluster-index of the selected
user during the $m$-th round.
\item Update: After the $m$-th round selection, the users in the $\mu\left(m\right)$
cluster are informed to keep silent afterwards, i.e., setting $\mathcal{K}\left(m+{\rm I}\right)=\mathcal{K}\left(m\right)\setminus\mu\left(m\right)$
and $m=m+{\rm I}$.
\item End control: If $m\leq\mathrm{III}$, go to step 2; else, end.
\end{enumerate}
After the selections, the {\color{black}indices} of the selected
users are collected in ${J_{D}}=\left\{ j_{\mu\left(1\right)}^{\ddagger},j_{\mu\left(2\right)}^{\ddagger},j_{\mu\left(3\right)}^{\ddagger}\right\} $.
As shown in Fig. 3, the proposed RSS-based scheduling is able to generate
the favorable patterns of characteristic directions, i.e., ${\left\{ {{\bf {r}}_{{j_{k}}}}\right\} _{k=1}^{3}}$
are respectively aligned with distinct directions of $\left\{ \mathbf{{e}}_{m}\right\} _{m={\rm I}}^{{\rm III}}$.
Then the channels as well as the SSA-resultant signal space are shaped
under the guidance of RSS.

Next, the detailed implementations of the 2) and 3) steps of the proposed
distributed CS are further elaborated. It is assumed that all the
users are synchronized to a common clock, such as the Global Positioning
System (GPS) signal, and a timer that lasts proportionally to $\phi_{j_{k},m}$
is installed in $\mathsf{S}_{j_{k}}$. {\color{black}More specifically,
with the clock period $T$, the response time of the timer of $\mathsf{S}_{j_{k}}$,
i.e., $\delta_{{j_{k}},m}$, can be defined as $\delta_{{j_{k}},m}=\frac{\phi_{{j_{k}},m}}{90^{o}}T$.
During the $m$-th round, $\mathsf{R}$ broadcasts a beacon signal
with the index of the successful candidate of the $\left(m-1\right)$-th
round, i.e., $j_{\mu\left(m-1\right)}^{\ddagger}$, and $j_{\mu\left(0\right)}^{\ddagger}=\mathrm{null}$.
Upon receiving the beacon, each user first checks if $\mu\left(m-1\right)$
equals to its own cluster-index. If so, the user keeps silent for
the rest of the scheduling period; if not, the user tries to compete
during this round. The competing user $\mathsf{S}_{j_{k}}$ obtains
the value of $m$ from its own counter, {\color{black}and calculates $\delta_{{j_{k}},m}$
to trigger the timer for the response (or the user's beacon) to the relay's beacon.} The first time-out
user is selected in each round.}

{\color{black}It is assumed that the transmission and processing times for different users are same and are smaller than the clock period $T$. Therefore, the transmission and processing times will not influence the scheduling decision. Moreover, the clock period $T$ is not too long and can be decided by hardware specification \cite{bletsas2006}. Therefore, the influence of $T$ on the system performance is negligible.}

\emph{Remark} 1: {\color{black}The distributed CS enjoys very low implementation
complexity without explicit feedback of CSI from the users.} Although a relatively
large number of candidates {\color{black}are} necessary to achieve
a distinct performance gain, the proposed scheme still enjoys a good
performance-complexity tradeoff for the considered scenario. It is
noted that the analysis of the number of candidates and the achievable
performance gain is very challenging. The obstacle is the unknown
statistical behaviour of the SSA-result signal space $\mathrm{\mathbf{F}}$
when user scheduling is considered \cite{GaoICC}. It will be {\color{black}shown}
later that the system overhead is fixed for the proposed distributed
CS, which is independent to the number of candidate users. In contrast,
the complexity and CSI overheads of the centralized CS increase very
fast as the number of users increases.

\subsubsection{RSS-based distributed GS}

Similar to the aforementioned CS, the proposed distributed GS relies
on the angular-coordinate calculated by each user with RSS and local
CSI. It is noted that the GS always {\color{black}needs} a certain
metric to evaluate the \emph{group performance}, which requires the
\emph{centralized decision}. Still, the proposed scheme aims to\emph{
distribute} \emph{the computations} to the users, so that the scheduling
center $\mathsf{R}$ can be designed as simple as possible. In particular,
the RSS-based distributed GS employs a progressive feedback protocol,
which consists of two phases. In the first phase, $\mathsf{S}_{j'_{k}}$
uses local CSI to check which direction of RSS is mostly aligned with
its characteristic direction $\mathbf{r}_{j'_{k}}$, and feeds back
the index of the most aligned RSS direction (using only two bits)
to R as
{\small\begin{equation}
m_{j'_{k}}=\arg\,\min_{m\in\left\{ {\rm I},{\rm II},{\rm III}\right\} }\,(\phi_{j'_{k},m}).\label{eq:pre_angle}
\end{equation}}Then $\mathsf{R}$ collects the {\color{black}indices} of each group
$J'$ in a set $M_{J'}=\left\{ m_{j'_{1}},m_{j'_{2}},m_{j'_{3}}\right\} $,
and check if the elements in $M_{J'}$ have distinctive values. This
checking serves as a coarse judgment on the orthogonality of the SSA-resultant
signal space $\mathbf{{\mathbf{F}}}$, and only when the elements
in $M_{J'}$ are of distinctive values the related users continue
to compete the channel. To offer some intuitions, Fig. 3 offers two
user combinations, namely $\left\{ {{\mathsf{S}_{{1_{1}}}},{\mathsf{S}_{{1_{2}}}},{\mathsf{S}_{{1_{3}}}}}\right\} $
and $\left\{ {{\mathsf{S}_{{2_{1}}}},{\mathsf{S}_{{2_{2}}}},{\mathsf{S}_{{2_{3}}}}}\right\} $.
The first combination can pass the coarse selection; but the second
can not, since both ${{\bf {r}}_{{2_{1}}}}$ and ${{\bf {r}}_{{2_{2}}}}$
are more aligned with ${{\bf {e}}_{{\rm I}}}$. Let us collect the
surviving groups in a set $\mathcal{J}''\subseteq\mathcal{J}'$, in
the second phase, $\mathsf{R}$ informs each surviving user to feed
back the individual scheduling metric, which is the {\color{black}smallest}
angle of its angular-coordinate within RSS, i.e.,
{\small\begin{equation}
\phi_{j'_{k},min}=\min_{m\in\left\{ {\rm I},{\rm II},{\rm III}\right\} }\,(\phi_{j'_{k},m}),\,{j'_{k}}\in{J'}\in{\cal J}''.\label{eq:feedback}
\end{equation}}Then $\mathsf{R}$ synthesizes $\left\{ \phi_{j'_{k},min}\right\} _{k=1}^{3}$
to generate the GS scheduling metric $\phi_{sum}^{\left(J'\right)}=\sum_{j'_{k}\in J'}\phi_{j'_{k},min}$
of one surviving group $J'\in{\cal J}''$, and the preferred user
group is selected as
{\small\begin{equation}
J'_{D}=\left\{ j_{1}^{'\ddagger},j_{2}^{'\ddagger},j_{3}^{'\ddagger}\right\} =\arg\,\min_{J'\in\mathcal{J}''}\,\left(\phi_{sum}^{\left(J'\right)}\right).\label{eq:d_GS}
\end{equation}}Finally, it is {\color{black}noted} that if no surviving users exist
after the first phase, random selection can be used to pick one group
out of ${\cal J}'$.

\emph{Remark} 2: The progressive feedback is an opportunistic feedback
scheme, which enables the distributed GS with low system overheads. {\color{black}In particular, in the second step of the protocol, the survived user feeds back the smallest angle in (\ref{eq:feedback}), which is a real number and requires the analog feedback. In practice, such analog feedback often requires quantization, but the detailed study on such implementation is beyond the scope of this paper and is left as our future work.}
Also noting the symmetry of the channel's statistic properties, it is easy
to check that the characteristic directions have equal chances to
align with every direction in $\mathbf{E}$ of RSS, then the \emph{average
surviving ratio} of a candidate group can be calculated as $3!/3^{3}=2/9$
after the first round of feedback on the aligned direction within
RSS (\ref{eq:pre_angle}). Therefore, only 2/9 of the user groups
need feed back the scheduling metrics (\ref{eq:feedback}) on average.
In this sense, the proposed GS is suitable for the networks where
the candidates are abundant and the low-complexity scheduling is demanded.

\emph{Remark} 3: In fact, the proposed distributed CS/GS can only
\emph{shape} the SSA-result signal space $\mathbf{{\mathbf{F}}}$,
which is not a straightforward optimization towards the post-processing
SNR. Since $\mathbf{{\mathbf{F}}}$ is coupled with the three-party
channels within a user group, any further descriptions on $\mathbf{{\mathbf{F}}}$
may require the three-party CSI. To this end, it seems that shaping
$\mathbf{{\mathbf{F}}}$ is perhaps the best thing one can do with
local CSI and RSS. It is noted that the statistical {\color{black}behavior}
of $\mathbf{{\mathbf{F}}}$ is unknown with user scheduling in general\cite{GaoICC},
and also because of the reasons explained by Remark 4 below, we do
not perform theoretical analysis on CS and GS with Min-UA transmission
in this paper.

\emph{Remark} 4: Similar to the distributed CS, the distributed GS
with Min-UA transmission can only harvest partial MuD gain when the
number of candidate groups is large; and the performance gap between
the centralized and distributed scheduling schemes are distinct, which
will be shown later in Section VI. These observations motivate us
to look into the ER-UA case, where near-optimal distributed scheduling
is possible as shown in the following sections.

{\color{black}

\subsection{RSS-based Distributed scheduling with Min-UA for $N>1$}

}

{\color{black}With a slight modification, the proposed distributed
CS and GS with Min-UA can be extended to a more general system model,
where each $2N$-antenna user transmits $2N$ data streams via a $3N$-antenna
$\mathsf{R}$, $N>1$. Following the same idea in  Subsection III-C, we choose the user
group to shape $\mathbf{F}$ with the help of RSS $\mathbf{E}$ and
local CSI. Specifically, the characteristic subspace instead of characteristic
direction of user $\mathsf{S}_{j_{k}}$ is introduced as $\mathbf{R}_{j_{k}}\in\textrm{Null}\left(\mathbf{H}_{j_{k}}\right)$,
where $\mathbf{R}_{j_{k}}\in\mathbb{C}^{3N\times N}$ and $\mathbf{R}_{j_{k}}^{H}\mathbf{R}_{j_{k}}=\mathbf{I}_{N\times N}$.
Furthermore, the chordal distance $d_{c}\left(\mathbf{A},\mathbf{B}\right)$
is used as an orthogonality measure of the two subspaces $\mathbf{A}$
and $\mathbf{B}$, where $d_{c}\left(\mathbf{A},\mathbf{B}\right)=\sqrt{N_T-\textrm{trace}\left(\mathbf{A}\mathbf{A}^{H}\mathbf{B}\mathbf{B}^{H}\right)}$
and $\mathbf{A},\mathbf{B}\in\mathbb{C}^{N_R\times N_T}$ are generator
matrices satisfying $\mathbf{A}^{H}\mathbf{A}=\mathbf{B}^{H}\mathbf{B}=\mathbf{I}_{N_T\times N_T}$\cite{golub2012matrix}.
The larger value of $d_{c}\left(\mathbf{A},\mathbf{B}\right)$ represents
the better orthogonality of $\mathbf{A}$ and $\mathbf{B}$. For the
procedure of distributed CS or GS with $N>1$, the angular-coordinate
of $\mathsf{S}_{j_{k}}$ is replaced by the chordal distance coordinate,
i.e., $\phi_{j_{k}}=[\phi_{j_{k},{\rm I}}\,\phi_{j_{k},{\rm II}}\,\phi_{j_{k},{\rm III}}]$,
where $\phi_{j_{k},m}=d_{c}\left(\mathbf{R}_{j_{k}},\mathbf{E}_{m}\right)$. Then, the distributed CS and GS can be carried our for $N>1$ with the newly defined $\phi_{j_{k}}$ and the same procedures in the previous subsections.

\emph{Remark} 5: It is worth pointing out that, in the scenario of $N>1$, the performance improvement
of the distributed CS with Min-UA transmission is not obvious
as the number of candidate increases, especially for
the distributed GS, which will be shown later in Section VI. In this scenario with high-dimensional signal space, it is very difficult to precisely characterize the orthogonality between subspaces
or the intersection spaces with local CSI, which limits the performance of
the proposed distributed scheduling schemes. Designing more effective
low-complexity scheduling schemes for Min-UA transmission with $N>1$ will be an interesting topic for future
research.}\\

%%%%%%%%%%%%%%%%%%%%%%%%%%%%%%%%%%%%%%%%%%%%%%%%%%%%%%%%%%%%%%%%%%%%%%

%%%%%%%%%%%%%%%%%%%%%%%%%%%%%%%%%%%%%%%%%%%%%%%%%%%%%%%%%%%%

\section{Distributed Scheduling with ER-UA Transmission}

In this section, the distributed user scheduling schemes are proposed
for ER-UA transmission, where both relay and users are equally equipped
with $N_{T}=N_{R}=3N$ antennas. As compared with the Min-UA transmission,
$N$ extra antennas are added at the {\color{black}users}, which
offers enough dimensions to achieve an active signal alignment with
predefined direction. Specifically, by venturing $N$ extra antennas
to each user, the application of RSS can be extended to guide both
distributed beamforming and user scheduling. Moreover, it is interesting
to note that, unlike the scheduling with Min-UA transmission, the
distributed user scheduling schemes achieve comparable performances
as their centralized counterparts with ER-UA transmission. Again,
the transmission scheme is first introduced before presenting the
user scheduling schemes.

%vectors $\mathbf{v}_{l,k}\in\mathbb{C}^{3\times1}$ and%$\mathbf{v}_{k,l}\in\mathbb{C}^{3\times1}$ at $\mathsf{S}_{k}$ and%$\mathsf{S}_{l}$, $m=\pi\left(l,k\right)$, $l\text{\ensuremath{\in}}\mathcal{L}_{k}$,%i.e., Fig. 4 illustrates the basic idea of ER-UA transmission.

\subsection{ER-UA MIMO-Y Transmission}

Again, let us assume three users $\left\{ \mathsf{S}_{1},\,\mathsf{S}_{2},\,\mathsf{S}_{3}\right\} $
are randomly selected to exchange information. It is noted that the
ER-UA MIMO-Y transmission relies on the RSS, which enables each user
to design its transmit beamforming vectors with only local CSI and
achieve SSA at the relay. Specifically, aiming at the pair-wise signal
alignment in RSS $\Omega_{R}$ at $\mathsf{R}$, the reference direction
$\mathbf{e}_{m}^{[n]}$ is allocated to guide the pair-wise transmit
beamforming of the $n$-th data stream at $\mathsf{S}_{k}$ and $\mathsf{S}_{l}$,
where $m=\pi\left(l,k\right)$ and $n\in[1,N]$. Note that the RSS-guided
transmit beamforming vectors $\mathbf{v}_{l,k}^{\left[n\right]}$
and $\mathbf{v}_{k,l}^{\left[n\right]}$ can be solved separately
with local CSI as
{\small\begin{equation}
\mathbf{v}_{l,k}^{\left[n\right]}=\sqrt{P_{T}/2N}\left\langle \mathbf{H}_{k}^{-1}\mathbf{e}_{m}^{\left[n\right]}\right\rangle ,\mathbf{v}_{k,l}^{[n]}=\sqrt{P_{T}/2N}\left\langle \mathbf{H}_{l}^{-1}\mathbf{e}_{m}^{\left[n\right]}\right\rangle ,\label{eq:TB_RSS}
\end{equation}}and the power constraint is imposed as $\left\Vert \mathbf{v}_{l,k}^{\left[n\right]}\right\Vert ^{2}=\left\Vert \mathbf{v}_{k,l}^{[n]}\right\Vert ^{2}=P_{T}/2N$
per data stream. During the MAC phase, it is observed that SSA is
achieved under the RSS, i.e., $\mathrm{Span}\left(\mathbf{H}_{k}\mathbf{v}_{l,k}^{\left[n\right]}\right)=\mathrm{Span}\left(\mathbf{H}_{l}\mathbf{v}_{k,l}^{[n]}\right)=\mathrm{Span}\left(\mathbf{e}_{m}^{\left[n\right]}\right)$,
as shown in Fig. 4 with $N=1$. the received signal at $\mathsf{R}$
(cf. (\ref{eq:M_Rx_Sig})) is given by
{\small\begin{align}
\mathbf{y}_{R} & =\sqrt{P_{T}/2N}\mathbf{E}\mathbf{\tilde{d}}_{+}+\mathbf{n}_{R},\label{eq:RSS_RX}
\end{align}}where $\mathbf{\tilde{d}}_{+}:=[\mathbf{\tilde{d}}_{+,\mathrm{I}}^{T}\,\mathbf{\tilde{d}}_{+,\mathrm{II}}^{T}\,\mathbf{\tilde{d}}_{+,\mathrm{III}}^{T}]^{T}\in\mathbb{C}^{3N\times1}$
is the vector of the superimposed signals and $\mathbf{E}=\left[\mathbf{E}_{{\rm I}}\,\mathbf{E}_{{\rm II}}\,\mathbf{E}_{{\rm III}}\right]\in\mathbb{C}^{N_{R}\times N_{R}}$
is the RSS as well as the the equivalent MIMO channel seen by $\mathsf{R}$.
The $m$-th component of $\mathbf{\tilde{d}}_{+}$ is $\mathbf{\tilde{d}}_{+,m}=\left[\tilde{d}_{+,m}^{[1]}\thinspace\tilde{d}_{+,m}^{[2]}\thinspace...\thinspace\tilde{d}_{+,m}^{[N]}\right]^{T}\in\mathbb{C}^{N\times1}$,
where $\tilde{d}_{+,m}^{\left[n\right]}=\tilde{d}_{l,k}^{[n]}+\tilde{d}_{k,l}^{[n]}$,
$\tilde{d}_{l,k}^{[n]}=\alpha_{m,k}^{[n]}d_{l,k}^{[n]}$, $m=\pi\left(l,k\right)$,
and $\alpha_{m,k}^{[n]}=\left\Vert \mathbf{H}_{k}^{-1}\mathbf{e}_{m}^{[n]}\right\Vert ^{-1}$
is the equivalent channel coefficient for $d_{l,k}^{[n]}$ in the
MAC phase. {\color{black}Specifically, the equivalent channel gain
(ECG) of $d_{l,k}^{[n]}$ is defined as $\left(\alpha_{m,k}^{[n]}\right)^{2}=\left\Vert \mathbf{H}_{k}^{-1}\mathbf{e}_{m}^{[n]}\right\Vert ^{-2}$.}
Aiming at a simpler implementation, the fixed-gain AF relay is used
here. The relay processing matrix is then simplified as
$\mathbf{\tilde{W}}=\mathbf{I}_{N_{R}\times N_{R}}$. During the BC
phase, the relay transmits $\mathbf{s}_{R}=G_{R}\mathbf{y}_{R}$ with
the long-term power controlling coefficient
{\small\begin{align}
G_{R} & =\sqrt{P_{R}/\mathrm{E}\left\{ \left\Vert \mathbf{y}_{R}\right\Vert ^{2}\right\} }=\sqrt{2NP_{R}/\left(P_{T}\bar{\alpha}_{sum}^{2}+2NN_{R}\sigma_{R}^{2}\right)}=\sqrt{P_{R}/3\left(P_{T}+N\sigma_{R}^{2}\right)},\label{eq:relay_gain}
\end{align}}where {\color{black}$\bar{\alpha}_{sum}^{2}=\sum_{k=1}^{3}\sum_{l\in\mathcal{L}_{k}}\sum_{n=1}^{N}{\left(\bar\alpha{}_{\pi\left(l,k\right),k}^{[n]}\right)^{2}}$=$6N$},
and ${\left(\bar\alpha{}_{m,k}^{[n]}\right)^{2}}=\mathrm{E}\left\{ \left\Vert \mathbf{H}_{k}^{-1}\mathbf{e}_{m}^{[n]}\right\Vert ^{-2}\right\} =1$
is the channel-averaged equivalent channel gain for $d_{l,k}^{[n]}$, which is calculated by employing the distribution of $\overline{\left(\alpha{}_{m,k}^{[n]}\right)^{2}}$ (see Lemma 4 in Appendix I). Note that $\overline{\left(\alpha{}_{m,k}^{[n]}\right)^{2}}$
is only related to the long-term channel statistics and can be calculated,
$G_{R}$ is thus treated as a constant and assumed to be known by
all the nodes in this network. Then the self-interference-free signal
$\tilde{\mathbf{y}}_{k}=\mathbf{y}_{k}-G_{R}\mathbf{H}_{k}^{H}\mathbf{I}\mathbf{H}_{k}\mathbf{s}_{k}$
at $\mathsf{S}_{k}$ is expanded as
{\small\begin{align}
\tilde{\mathbf{y}}_{k} & =\sqrt{P_{T}/2N}G_{R}\mathbf{H}_{k}^{H}\mathbf{\breve{E}}_{k}\mathbf{\breve{d}}_{+,k}+G_{R}\mathbf{H}_{k}^{H}\mathbf{n}_{R}+\mathbf{n}_{k},\label{eq:source}
\end{align}}where $\mathbf{\breve{E}}_{k}=\left[\mathbf{E}_{m_{1}}\,\mathbf{E}_{m_{2}}\,\mathbf{E}_{m'}\right]\in\mathbb{C}^{N_{R}\times N_{R}}$,
and $\mathbf{\breve{d}}_{+,k}=\left[\tilde{\mathbf{d}}_{k,l_{1}}\,\tilde{\mathbf{d}}_{k,l_{2}}\,\tilde{\mathbf{d}}_{+,m'}\right]^{T}\in\mathbb{C}^{N_{T}\times1}$,
$m_{1}=\pi\left(l_{1},k\right)$, $m_{2}=\pi\left(l_{2},k\right)$,
$l_{1},\, l_{2}\in\mathcal{L}_{k}$ and $m'\in\left\{ \mathrm{I},\mathrm{II},\mathrm{II}\right\} \backslash\left\{ m_{1},m_{2}\right\} .$
Based on (\ref{eq:source}), the receive beamforming $\mathbf{U}_{k}=G_{R}^{-1}\left[\mathbf{H}_{k}^{-1}\mathbf{E}_{m_{1}}\,\mathbf{H}_{k}^{-1}\mathbf{E}_{m_{2}}\right]\in\mathbb{C}^{N_{T}\times2N}$
is used to obtain $\mathbf{\hat{y}}_{k}=\mathbf{U}_{k}^{H}\tilde{\mathbf{y}}_{k}$,
and each desired signal stream within $\mathbf{\hat{y}}_{k}$ is given
by
{\small\begin{equation}
\hat{y}_{k,l}^{[n]}=\sqrt{P_{T}/2N}\alpha_{m,l}^{[n]}d_{k,l}^{[n]}+n_{k,l}^{[n]},\, l\text{\ensuremath{\in}}\mathcal{L}_{k},n\in[1,N],\label{eq:EQ-P2P}
\end{equation}}where $n_{k,l}^{[n]}=\left(\mathbf{e}_{m}^{[n]}\right)^{H}\mathbf{n}_{R}+\frac{1}{G_{R}}\left(\mathbf{e}_{m}^{[n]}\right)^{H}\left(\mathbf{H}_{k}^{-1}\right)^{H}\mathbf{n}_{k}$
is the noise item. Then it is easy to extract the useful information
from $\hat{y}_{k,l}^{[n]}$.
\begin{figure}[t]
\centering \includegraphics[width=9cm]{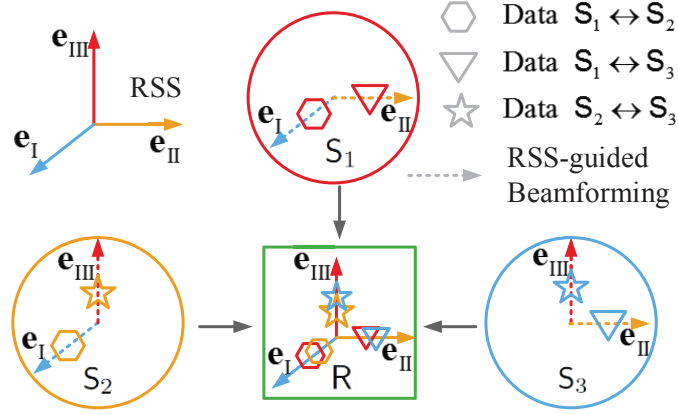} \protect\protect\caption{Geometrical interpretation on the ER-UA transmission. After the RSS-guided
beamforming, the signals are pair-wisely aligned with the basis of
RSS at the relay during the MAC phase. The aligned directions are
orthogonal, and the equivalent MIMO channel seen by the relay is well-conditioned.}
\end{figure}

\emph{Remark} 6: It is noted that the Min-UA transmission scheme requires\emph{
joint} transmit/receive beamforming design with the \emph{three-party}
CSI at users and relay, which involves high CSI overheads and relatively
complicated signal processing. In contrast to Min-UA, ER-UA transmission
employs extra user antennas to enable the simple RSS-based \emph{distributed}
transmit/receive beamforming design with \emph{local} CSI. Therefore,
the system overhead for CSI exchanging is significantly reduced. Another
advantage of ER-UA transmission is that it enables the near-optimal
and low-complexity RSS-based distributed user scheduling, which will
be shown in the next subsection.

\subsection{Problem Formulation and Centralized Scheduling}

Similar to Section III-B%
\footnote{We may reuse some of the notations appeared in the previous sections
to convey similar concepts, if not causing confusion.%
}, the user index set $J=\left\{ j_{1},j_{2},j_{3}\right\} $ is re-introduced
and the relevant overall outage probability is defined as $P_{out}^{\left(J\right)}\left(\rho_{th}\right)=\Pr\left(\rho_{min}^{\left(J\right)}\leq\rho_{th}\right)$
and the overall post-processing SNR is defined as $\rho_{min}^{\left(J\right)}=\min_{k\in[1,3],\, l\text{\ensuremath{\in}}\mathcal{L}_{k},\thinspace n\in[1,d]}\left(\rho_{j_{k},j_{l}}^{[n]}\right)$,
\textcolor{black}{where $\rho_{j_{k},j_{l}}^{[n]}$ is the end-to-end post-processing
SNR of the $n$-th data stream within the link $\mathsf{\mathsf{S}}_{j_{l}}\rightarrow\mathsf{R}\rightarrow\mathsf{\mathsf{S}}_{j_{k}},\, l\text{\ensuremath{\in}}\mathcal{L}_{k}$.
From (\ref{eq:EQ-P2P}), {\small$\ensuremath{\rho_{j_{k},j_{l}}^{[n]}}=\mathrm{\mathsf{E}}\left\{ \left(\sqrt{P_{T}/2N}\alpha_{m,l}^{[n]}d_{k,l}^{[n]}\right)^{2}\right\} /\mathsf{E}\left\{ \left(n_{k,l}^{[n]}\right)^{2}\right\}$} is calculated
as
{\small\begin{equation}
\rho_{j_{k},j_{l}}^{[n]}=\frac{1}{2N}\cdot\frac{\rho_{j_{k},j_{l},2}^{[n]}\rho_{j_{k},j_{l},1}^{[n]}}{\rho_{j_{k},j_{l},2}^{[n]}+3\left(\mathrm{SNR}_{T}+N\right)},\label{eq:Link_Pout}
\end{equation}}where $\rho_{j_{k},j_{l},1}^{[n]}=\mathrm{SNR}_{T}\left(\alpha_{m,j_{l}}^{[n]}\right)^{2}$
and $\rho_{j_{k},j_{l},2}=\mathrm{SNR}_{R}\left(\alpha_{m,j_{k}}^{[n]}\right)^{2}$
are treated as the equivalent SNRs of the first hop and second hop, {\color{black}$\mathrm{SNR}_{T}=P_T/\sigma_{R}^{2}$ and $\mathrm{SNR}_{R}=P_R/\sigma_{R}^{2}$}}.
After introducing necessary notations, the centralized CS and GS are
{\color{black}first} considered as benchmarks, which are respectively
given by
{\small\begin{equation}
J_{C}=\left\{ j_{1}^{*},j_{2}^{*},j_{3}^{*}\right\} =\arg\,\max_{J\in\mathcal{J}}\,\left(\rho_{min}^{\left(J\right)}\right),\label{eq:outage optimal}
\end{equation}}
and
{\small\begin{equation}
J'_{C}=\left\{ j_{1}^{'*},j_{2}^{'*},j_{3}^{'*}\right\} =\arg\,\max_{J'\in\mathcal{J}'}\,\left(\rho_{min}^{\left(J'\right)}\right).\label{eq:group-wise centralized}
\end{equation}}Again, it is noted that the CSI overheads and the computational complexities
involved in the centralized CS and GS are high. To this end, a simplified
scheduling is required.

%\begin{figure}[t]%\centering \includegraphics[width=8.5cm]{Illustration421} \caption{The geometrical interpretation on the SSA-resultant signal space with Min-UA transmission. $\mathbf{\tilde{e}}_{m=\pi\left(k,l\right)}$ is the joint space of ${{\bf{H}}_{{j_k}}}$ and ${{\bf{H}}_{{j_l}}}$, $k,l \in \left[ {1,3} \right],\;k \ne l$; ${{\bf{r}}_{{j_k}}}$ is the characteristic direction of the channel $\mathbf{H}_{j_{k}}$, defined as $\mathbf{r}_{j_{k}}\in\mathrm{Null}\left(\mathbf{H}_{j_{k}}\right)$,%$\left\Vert \mathbf{r}_{j_{k}}\right\Vert =1$. The angles between ${{{\bf{\tilde e}}}_{\pi \left( {l,n} \right)}}$ and ${{\bf{r}}_{{j_k}}}$, for all $l,n,k \in \left[ {1,3} \right],\;l \ne n \ne k$ reflect the orthogonality among channels as  well as the SSA-resultant signal space with Min-UA transmission.}%\end{figure}

\begin{figure}[t]
\centering \includegraphics[width=9cm]{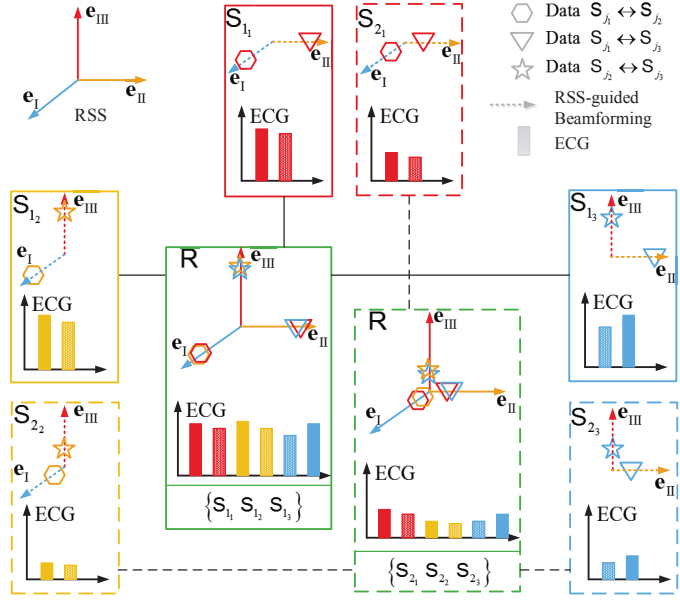} \protect\caption{The geometrical interpretations on the RSS-based distributed user
scheduling with ER-UA transmission. In this example, we consider the
cluster-wise scheduling and assume that there are two users $\left\{ {{\mathsf{S}_{{1_{k}}}},{\mathsf{S}_{{2_{k}}}}}\right\} $
in the $k$-th cluster, $k\in\left[{1,3}\right]$. {\color{black}Due}
to the RSS-guided transmit beamforming design, each user can predict
the ECG of its own signals before their arrival at the relay. Employing
the proposed distributed user scheduling, the minimum ECG at the relay
can be directly improved.}
\end{figure}

\subsection{RSS-based Distributed User Scheduling with ER-UA}

In this subsection, RSS-based distributed CS and GS are proposed with
ER-UA transmission. {\color{black} The main idea of
the RSS-based scheduling here is that, employing the RSS-based beamforming,
each user can perfectly align its signal space with the pre-defined
subspace of the RSS $\mathbf{E}$ (thanks to the ER-UA configuration),
and the user can further calculate its scheduling metric with local
CSI and inform the relay in a very efficient way. Unlike the Min-UA
scenario, the user scheduling metric in ER-UA scenario is directly
related to the end-to-end SNR, therefore, the effectiveness of the
proposed user scheduling is more prominent. More specifically, employing
the RSS-based beamforming the SSA-resultant signal space is same as
the RSS $\mathbf{E}$ due to perfect alignment. In addition, the effective
MIMO channel can be decoupled and the end-to-end link SNR of one data
stream in (\ref{eq:Link_Pout}) is an increasing function of the two
equivalent channel gains (ECGs) defined in Section IV-A,$\left(\alpha_{m,j_{l}}^{[n]}\right)^{2}$
and $\left(\alpha_{m,j_{k}}^{[n]}\right)^{2}$, where each ECG is
only determined by the local CSI of a user and the RSS. Based on this
observation, we propose disturbed user scheduling to maximize the
ECG as well as the end-to-end SNR, and the efficient implementations
are detailed in the following subsections.}

\subsubsection{RSS-based Distributed CS}

{\color{black}Employing RSS, each user can not only design its transmit
beamforming as (\ref{eq:TB_RSS}) to ensure SSA at the relay, but
also calculate the scheduling metric, i.e., the minimum-ECG, to enable
distributed CS. In particular, the minimum-ECG of $\mathsf{S}_{j_{k}}$
is defined as $\alpha_{j_{k}}^{2}=\min_{n\in[1,N]}\left(\min\,\left(\left(\alpha_{m_{1},j_{k}}^{[n]}\right)^{2},\left(\alpha_{m_{2},j_{k}}^{[n]}\right)^{2}\right)\right)$,
$m_{1}=\pi\left(l_{1},k\right)$, $m_{2}=\pi\left(l_{2},k\right)$,
and $\left(\alpha_{m_{1},j_{k}}^{[n]}\right)^{2}$ and $\left(\alpha_{m_{2},j_{k}}^{[n]}\right)^{2}$
are the ECGs, which are defined in Section IV A, for $d_{l_{1},k}^{[n]}$
and $d_{l_{2},k}^{[n]}$ separately. $\alpha_{j_{k}}^{2}$ can be
calculated with the local CSI $\mathbf{H}_{j_{k}}$ and RSS. It is
noted the RSS-based distributed CS is independently conducted in each
cluster, which aims to find the user with the maximal minimum-ECG
from each cluster, as shown in Fig. 5.} Then, the efficient distributed
scheduling is given by $J_{D}=\left\{ j_{1}^{\ddagger},j_{2}^{\ddagger},j_{3}^{\ddagger}\right\} $,
where the preferable user of the $k$-th cluster is selected according
to the following criterion
{\small\begin{align}
j_{k}^{\ddagger}=\arg\,\max_{j_{}\text{\ensuremath{\in}}\left[1,M_{k}\right]}\,\left(\alpha_{j_{k}}^{2}\right), k\in[1,3].\label{eq:d-UE}
\end{align}}The distributed implementation of the proposed scheme is simple. Similar
to distributed CS with Min-UA transmission, we assume all users are
synchronized to a common clock. {\color{black}To start the scheduling,
$\mathsf{R}$ broadcasts a beacon and $\mathsf{S}_{j_{k}}$ calculates
$\alpha_{j_{k}}^{2}$ with local CSI $\mathbf{H}_{j_{k}}$; then a
timer that lasts inverse-proportionally to $\alpha_{j_{k}}^{2}$ is
used by $\mathsf{S}_{j_{k}}$. Specifically, with the clock period
$T$, the response time of the timer of $\mathsf{S}_{j_{k}}$, i.e.,
$\delta'_{{j_{k}}}$ can be defined as $\delta'_{{j_{k}}}=\frac{1}{\alpha_{j_{k}}^{2}}T$.
Then, the competing user $\mathsf{S}_{j_{k}}$ calculates $\delta_{{j_{k}}}$
to trigger the timer for the response to the beacon. The first time-out
user must be $\mathsf{S}_{j_{k}^{\ddagger}}$.}

\subsubsection{RSS-based Distributed GS}

Similar to the Min-UA scenario, for the distributed GS with ER-UA
transmission, each user exploits local CSI to calculate the individual
scheduling metric and feeds it back to $\mathsf{R}$ for final decision.
The distributed GS is first given by
{\small\begin{equation}
J'_{D}=\left\{ j_{1}^{'\ddagger},j_{2}^{'\ddagger},j_{3}^{'\ddagger}\right\} =\arg\,\max_{J'\in\mathcal{J}'}\,\left(\gamma_{min}^{\left(J'\right)}\right),\label{eq:group-wise distributed}
\end{equation}}where $\gamma_{min}^{\left(J'\right)}=\frac{1}{2N}\cdot\frac{\mathrm{SNR}_{T}\left(\alpha_{[3]}^{\left(J'\right)}\right)^{2}\mathrm{SNR}_{R}\left(\alpha_{[2]}^{\left(J'\right)}\right)^{2}}{\mathrm{SNR}_{R}\left(\alpha_{[2]}^{\left(J'\right)}\right)^{2}+3\left(\mathrm{SNR}_{T}+N\right)}$
is the GS metric synthesized by $\mathsf{R}$ and is defined as the
equivalent SNR of the user group $J'$, and $\alpha_{[n]}^{\left(J'\right)}$\emph{
}is the $n$-th largest element of\emph{$\left\{ \alpha_{j'_{1}},\,\alpha_{j'_{2}},\,\alpha_{j'_{3}}\right\} $}.
It is noted that user $\mathsf{S}_{j_{k}}$ can calculate $\alpha_{j'_{k}}^{2}$
with local CSI, and the value is fed back to the relay $\mathsf{R}$.
With a total of $3M$ feedback of individual metrics, $\mathsf{R}$
forms the set $\left\{ \gamma_{min}^{\left(J'_{p}\right)}\right\} _{p=1}^{M}$
and makes a centralized decision to choose the preferred user group.

\emph{Remark} 7: The individual scheduling metric and the synthesized
GS metric are critical. In the proposed distributed GS, $\alpha_{j_{k}}^{2}$
characterizes the quality of the weaker link between $\mathsf{\mathsf{S}}_{k}$
and $\mathsf{R}$, and $\gamma_{min}^{\left(J'\right)}$ is actually
constructed as a lower bound of $\rho_{min}^{\left(J'\right)}$ to
be shown later. Therefore, the distributed GS aims to improve the
lower bound of the overall system performance.

\emph{Remark} 8: Unlike the scheduling schemes with the Min-UA transmission,
the proposed distributed CS and GS achieve comparable performances
as their centralized counterparts with the ER-UA; therefore, the proposed
distributed scheduling schemes enjoy very good performance-complexity
tradeoffs with the ER-UA. In the next section, these observations
are theoretically analyzed.

{\color{black}\emph{Remark} 9: For Min-UA transmission, the scheduling metrics of
the distributed CS and GS are determined by the angular-coordinate
$\phi_{j_{k}}$. Due to the symmetrical random property of the considered
wireless channel, the distribution of $\phi_{j_{k}}$ is identical for different $j_{k}$; therefore, each user or user group have the same opportunity
to be selected on the long-term. For ER-UA transmission, the scheduling metrics of the
distributed CS and GS are determined by the minimum-ECG $\alpha_{j_{k}}$. It is easy to check that for the considered scenarios, the distribution of $\alpha_{j_{k}}$ is identical for different $j_{k}$; therefore, the the long-term fairness
can also be guaranteed.}

\section{Performance Analysis with ER-UA Transmission}

%For $\rho_{min,CS}^{\left(J_{C}\right)}$ and $\rho_{min,GS}^{\left(J_{D}\right)}$,%i.e., the overall post-processing SNRs with the centralized $\left({J_{C}}\right)$%and distributed $\left({J_{D}}\right)$ scheduling;

In this section, outage performances of the proposed distributed CS,
GS and their centralized counterparts are quantified with ER-UA transmission.
Since we are mainly interested in the MuD orders, we assume $N=1$,
$\mathrm{SNR}_{R}=\mathrm{SNR}_{T}=\mathrm{SNR}$ and $M_{1}\leq M_{2}\leq M_{3}$
to ease the derivations. {\color{black}Note that $N=1$, the RSS and the ECG are simplified as $\mathbf{E}=\left[\mathbf{e}_{{\rm I}}\,\mathbf{e}_{{\rm II}}\,\mathbf{e}_{{\rm III}}\right]\in\mathbb{C}^{3\times3}$
and $\alpha_{m,j_{k}}^{2}=\left\Vert \mathbf{H}_{j_{k}}^{-1}\mathbf{e}_{m}\right\Vert ^{-2}$,
and $\gamma_{min}^{\left(J'\right)}$ in (\ref{eq:group-wise distributed}) is simplified as $\gamma_{min}^{\left(J'\right)}=\frac{\left(\alpha_{[3]}^{\left(J'\right)}\right)^{2}\left(\alpha_{[2]}^{\left(J'\right)}\right)^{2}\mathrm{SNR}}{2\left(\alpha_{[2]}^{\left(J'\right)}\right)^{2}+6\left(1+\mathrm{SNR}^{-1}\right)}$. In order to facilitate the analysis, we abstract the structure of $\gamma_{min}^{\left(J'\right)}$
as a function $g\left(x,y\right)=\frac{xy\mathrm{SNR}}{2y+6\left(1+\mathrm{SNR}^{-1}\right)}$ for wider applications}, and we also denote the min-SNRs (cf. (\ref{eq:Pout})) of centralized CS (\ref{eq:outage optimal})
and GS (\ref{eq:group-wise centralized}) as $\rho_{min,CS}^{\left(J_{C}\right)}$
and $\rho_{min,GS}^{\left(J'_{C}\right)}$; similarly, we denote the
min-SNRs of distributed CS (\ref{eq:d-UE}) and GS (\ref{eq:group-wise distributed})
as $\rho_{min,CS}^{\left(J_{D}\right)}$ and $\rho_{min,GS}^{\left(J'_{D}\right)}$.
It is noted that the analysis {\color{black}of} the distributed
scheduling schemes {\color{black}is} difficult, because the optimization
objectives of these schemes are not exactly the min-SNRs of the network.
To this end, tractable lower {\color{black}bounds} (LB) and upper
{\color{black}bounds} (UB) are first established for 1) $\rho_{min,CS}^{\left(J_{C}\right)}$
and $\rho_{min,CS}^{\left(J_{D}\right)}$ with CS, then for 2) $\rho_{min,GS}^{\left(J'_{C}\right)}$
and $\rho_{min,GS}^{\left(J'_{D}\right)}$ with GS, respectively.
Next, the bounds of outage probabilities are developed with tractable
theoretical results, and their high SNR approximations are analyzed
to extract the achievable MuD orders.

\subsection{Bounding the outage probabilities }

%\begin{align}%I_{1,2} & =\frac{M_{2}\left(1-e^{-3\mu}\right)^{M_{1}+M_{2}}\left[1-\left(1-e^{-3\mu}\right)^{M_{3}}\right]}{M_{1}+M_{2}}.\label{eq:I2}%\end{align}

To begin with, the following proposition is introduced to bound the
min-SNRs with CS.
\begin{prop}
\emph{Using ER-UA transmission, the min-SNRs of centralized and distributed
CS are bounded as}
{\small\begin{equation}
{\color{black}\rho_{min,CS}^{LB}}\overset{\left(a1\right)}{\leq}\rho_{min,CS}^{\left(J_{D}\right)}\overset{\left(a2\right)}{\leq}\rho_{min,CS}^{\left(J_{C}\right)}\overset{\left(a3\right)}{\leq}{\color{black}\rho_{min,CS}^{UB}},\label{eq:the bound}
\end{equation}}\emph{where the LB and UB are ${\color{black}\rho_{min,CS}^{LB}}=g\left(\tilde{\lambda}_{[3]}^{\left(J_{\lambda}\right)},\tilde{\lambda}_{[2]}^{\left(J_{\lambda}\right)}\right)$ and ${\color{black}\rho_{min,CS}^{UB}}=g\left(\alpha_{\mathrm{I},j_{1}^{\dagger}}^{2},\alpha_{\mathrm{I},j_{2}^{\dagger}}^{2}\right)$,
respectively, }\textup{ and }\emph{$g\left(x,y\right):=\frac{xy\mathrm{SNR}}{2y+6\left(1+\mathrm{SNR}^{-1}\right)}$,
$x,y>0$. For ${\color{black}\rho_{min,CS}^{LB}}$, $\tilde{\lambda}_{[n]}^{\left(J_{\lambda}\right)}$
is the $n$-th largest element of $\left\{ \tilde{\lambda}_{j_{1}^{\star}},\,\tilde{\lambda}_{j_{2}^{\star}},\,\tilde{\lambda}_{j_{3}^{\star}}\right\} ,$
where $J_{\lambda}=\left\{ j_{1}^{\star},j_{2}^{\star},j_{3}^{\star}\right\} $,
}\textup{$j_{k}^{\star}=\arg\,\max_{j_{k}\in[1,M_{k}]}\left(\tilde{\lambda}_{j_{k}}\right)$
and}\emph{ $\tilde{\lambda}_{j_{k}}=\lambda_{min}\left(\mathbf{H}_{j_{k}}\mathbf{H}_{j_{k}}^{H}\right),$
$j_{}\in[1,M_{k}]$, $k\in[1,3]$. For ${\color{black}\rho_{min,CS}^{UB}}$, $j_{k}^{\dagger}=\arg\,\max_{j_{}\in[1,M_{k}]}\left(\alpha_{\mathrm{I},j_{k}}^{2}\right)$,
$k=1,2$. }\end{prop}
\begin{IEEEproof}
See Appendix I and II.
\end{IEEEproof}
Similarly, the following proposition is used to bound the min-SNRs
with GS.
\begin{prop}
\emph{Using ER-UA transmission, the min-SNRs of centralized GS and
distributed GS are bounded as}
{\small\begin{equation}
{\color{black}\rho_{min,GS}^{LB}}\overset{\left(b1\right)}{\leq}\rho_{min,GS}^{\left(J'_{D}\right)}\overset{\left(b2\right)}{\leq}\rho_{min,GS}^{\left(J'_{C}\right)}\overset{\left(b3\right)}{\leq}{\color{black}\rho_{min,GS}^{UB}},\label{eq:the bound-1}
\end{equation}}\emph{where the LB and UB are} \textup{${\color{black}\rho_{min,GS}^{LB}}=\max_{J'\in\mathcal{J}'}\left\{ g\left(\tilde{\lambda}_{[3]}^{\left(J'\right)},\tilde{\lambda}_{[2]}^{\left(J'\right)}\right)\right\} $
and ${\color{black}\rho_{min,GS}^{UB}}=g\left(\alpha_{\mathrm{I},j_{1}^{'\dagger}}^{2},\alpha_{\mathrm{I},j_{2}^{'\dagger}}^{2}\right)$,
respectively. For ${\color{black}\rho_{min,GS}^{LB}}$,} $\tilde{\lambda}_{[n]}^{\left(J'\right)}$
\emph{is the $n$-th largest element of }$\left\{ \tilde{\lambda}_{j'_{1}},\,\tilde{\lambda}_{j'_{2}},\,\tilde{\lambda}_{j'_{3}}\right\} ,$
\emph{where} $J'=\left\{ j'_{1},j'_{2},j'_{3}\right\} \in\mathcal{J}'$,
$\tilde{\lambda}_{j'_{k}}=\lambda_{min}\left(\mathbf{H}_{j'_{k}}\mathbf{H}_{j'_{k}}^{H}\right)$\emph{.
For} \textup{${\color{black}\rho_{min,GS}^{UB}}$, }\emph{$j_{k}^{\dagger}=\arg\,\max_{j_{}\in[1,M]}\left(\alpha_{\mathrm{I},j_{k}}^{2}\right)$,
$k=1,2$. }\end{prop}
\begin{IEEEproof}
See Appendix I and III.
\end{IEEEproof}
From Proposition 1 and Proposition 2, the \emph{common} LB and UB
for $P_{out,CS}^{\left(J_{C}\right)}$ and $P_{out,CS}^{\left(J_{D}\right)}$
with CS are defined as
{\small\begin{equation}
P_{out,CS}^{LB}\leq P_{out,CS}^{\left(J_{C}\right)}\leq P_{out,CS}^{\left(J_{D}\right)}\leq P_{out,CS}^{UB},\label{eq:CC_Bound}
\end{equation}}where $P_{out,CS}^{LB}\left(\rho_{th}\right)=\Pr\left({\color{black}\rho_{min,CS}^{UB}}\leq\rho_{th}\right),$
$P_{out,CS}^{UB}\left(\rho_{th}\right)=\Pr\left({\color{black}\rho_{min,CS}^{LB}}\leq\rho_{th}\right)$.
And similarly, the common LB and UB for $P_{out,GS}^{\left(J'_{C}\right)}$
and $P_{out,GS}^{\left(J'_{D}\right)}$ with GS are given by
{\small\begin{equation}
P_{out,GS}^{LB}\leq P_{out,GS}^{\left(J'_{C}\right)}\leq P_{out,GS}^{\left(J'_{D}\right)}\leq P_{out,GS}^{UB},\label{eq:GC_Bound}
\end{equation}}where $P_{out,GS}^{LB}\left(\rho_{th}\right)=\Pr\left({\color{black}\rho_{min,GS}^{UB}}\leq\rho_{th}\right),$
$P_{out,GS}^{UB}\left(\rho_{th}\right)=\Pr\left({\color{black}\rho_{min,GS}^{LB}}\leq\rho_{th}\right)$.
Then, these bounds are further quantified. Particularly, an ordered
set $\mathcal{M}_{\pi}=\left\{ \mathcal{M}_{i}\right\} _{i=1}^{6}$
is introduced to collect all the permutations of the three elements
in {\color{black}$\left\{ M_{1},\, M_{2},\, M_{3}\right\} $},
where $\mathcal{M}_{i}$ is also an ordered set and the $n$-th element
of $\mathcal{M}_{i}$ is denoted as $M_{i,n}$, $n\in[1,3]$. The
detailed derivations are collected in Appendix IV and the key results
of $P_{out,CS}^{LB}\left(\rho_{th}\right)$, $P_{out,CS}^{UB}\left(\rho_{th}\right)$
and $P_{out,GS}^{UB}\left(\rho_{th}\right)$ are given by {\small{}
\begin{equation}
P_{out,CS(GS)}^{LB}\left(\rho_{th}\right)=2M_{2}\sum_{q=0}^{M_{2}-1}\left[\frac{\left(-1\right)^{q}{\color{black}\tbinom{M_{2}-1}{q}}}{2\left(q+1\right)}+\sum_{p=1}^{M_{1}}{\color{black}\tbinom{M_{2}-1}{q}}{\color{black}\tbinom{M_{1}}{p}}\left(-1\right)^{q+p}e^{-pa}\sqrt{\frac{pb}{\left(q+1\right)}}K_{1}\left(2\sqrt{p\left(q+1\right)b}\right)\right],\label{eq:LB}
\end{equation}
} {\small{}
\begin{align}
P_{out,CS}^{UB}\left(\rho_{th}\right) & \approx\sum_{i=1}^{6}\left(\frac{M_{i,2}M_{i,3}\left(1-e^{-3\mu}\right)^{M_{\Sigma}}+M_{i,2}M_{\Sigma}\left(1-e^{-3\mu}\right)^{M_{i,1}+M_{i,2}}\left[1-\left(1-e^{-3\mu}\right)^{M_{i,3}}\right]}{\left(M_{i,1}+M_{i,2}\right)M_{\Sigma}}\right.\label{eq:UB}\\
 & \left.+\left[3M_{i,2}\sum_{p=0}^{M_{i,2}-1}\sum_{q=0}^{M_{i,1}}\left(-1\right)^{q}{\color{black}\tbinom{M_{i,2}-1}{p}}{\color{black}\tbinom{M_{i,1}}{q}}e^{-3qa}\int_{\mu}^{\infty}e^{-3(\left(p+1\right)y-qb\frac{1}{y})}dy\right]\left[1-\left(1-e^{-3\mu}\right)^{M_{i,3}}\right]\right),\nonumber
\end{align}
} and {\small{}
\begin{equation}
P_{out,GS}^{UB}\left(\rho_{th}\right)\approx\left[3\left(1-e^{-3\mu}\right)^{2}-2\left(1-e^{-3\mu}\right)^{3}+6\left(e^{-6\mu}-3e^{-3\left(\mu+a\right)}\int_{\mu}^{\infty}e^{-3[y-b\frac{1}{y}]}dy\right)\right]^{M},\label{eq:UB1}
\end{equation}
}respectively, where $K_{1}\left(x\right)$ is the modified Bessel
function of the second kind \cite{Table}, $a\left(\rho_{th}\right)=\frac{2\rho_{th}}{\mathrm{SNR}}$,
$b\left(\rho_{th}\right)=\frac{6\rho_{th}\left(\mathrm{SNR}+1\right)}{\mathrm{SNR^{2}}}$
and $\mu=\frac{1}{2}\left(a+\sqrt{a^{2}+4b}\right)$ is the positive
root of the quadratic equation $y^{2}=ay+b$, $M_{\Sigma}=M_{1}+M_{2}+M_{3}$.
It is noted that the integral $\int_{\mu}^{\infty}e^{-3[\left(p+1\right)y-qb\frac{1}{y}]}dy$
in (\ref{eq:UB}) can be efficiently evaluated with software like
MATLAB or Mathematica. Finally, it is noted that $P_{out,GS}^{LB}\left(\rho_{th}\right)$
can be obtained from (\ref{eq:LB}) by setting $M_{1}=M_{2}=M$, therefore,
the equation of (\ref{eq:LB}) is reused for conciseness.

\subsection{High SNR Analysis}

In this subsection, the high SNR analysis on the bounds of outage
probabilities are given. Only the key results are provided here while all the standard derivations
and are collected in Appendix V and Appendix VI.

\subsubsection{High SNR Approximations of \textmd{$P_{out,CS}^{LB}$ and $P_{out,CS}^{UB}$}}

\paragraph{$P_{out,CS}^{LB}$}

%Recall the $n$-th largest element of $\mathcal{M}$ is denoted as $M_{[n]}$,%thenThe
high SNR approximation of $P_{out,CS}^{LB}\left(\rho_{th}\right)$
is given as follows {{
{\small\begin{equation}
\left\{ \begin{array}{cc}
G_{CS,1}^{LB}\left(\frac{2\rho_{th}}{\mathrm{SNR}}\right)^{d_{CS}^{UB}}\ln\left(\frac{\mathrm{SNR}}{2\rho_{th}}\right), & M_{[2]}=M_{[3]},\\
G_{CS,2}^{LB}\left(\frac{2\rho_{th}}{\mathrm{SNR}}\right)^{d_{CS}^{UB}}, & M_{[2]}\neq M_{[3]},
\end{array}\right.\label{eq:LB_High}
\end{equation}}}}where the diversity UB is $d_{CS}^{UB}=M_{[3]}$, and the power
gains are $G_{CS,1}^{LB}=M_{[3]}3^{M_{[3]}},$ $G_{CS,2}^{LB}=\varphi\left(M_{[3]},M_{[2]},1\right)$.
Here $M_{[n]}$ is the $n$-th largest element of $\left\{ M_{1},M_{2},M_{3}\right\} $,
and $\varphi\left(N_{1},N_{2},\tau\right)$ is given by{\small{}{
\[
\varphi\left(N_{1},N_{2},\tau\right)=N_{2}\sum_{q=0}^{N_{2}-1}\sum_{p=1}^{N_{1}}\frac{\left(-1\right)^{q+p}{\color{black}\tbinom{N_{2}-1}{q}}{\color{black}\tbinom{N_{1}}{p}}\left(\mathbf{c}_{p}^{T}(\tau)\mathbf{e}_{p,q}\left(\tau\right)\right)}{\left(q+1\right)},
\]
}}where $\mathbf{c}_{p}\left(\tau\right)=[c_{p,d'}\left(\tau\right),\, c_{p,d'-1}\left(\tau\right),\ldots,\,1]^{T}\in\mathbb{C}^{(d'+1)\times1}$
with element $c_{p,n}(\tau)=\frac{\left(-\tau p\right)^{n}}{n!}$,
$n\in[0,d']$, $d'=\min\left(N_{1},N_{2}\right)$, and $\mathbf{e}_{p,q}\left(\tau\right)=\left[0,\, e_{p,q,1}\left(\tau\right),\ldots,\, e_{p,q,d'}\left(\tau\right)\right]^{T}\in\mathbb{C}^{(d'+1)\times1}$
with element {\small{}{
\begin{align*}
e_{p,q,n}\left(\tau\right) & =\left(b_{1,n-1}\ln\left(2\tau\sqrt{3p\left(q+1\right)}\right)+b_{2,n-1}\right)\\
 & \qquad\times\left(12\tau^{2}p\left(q+1\right)\right){}^{n},\, n\in[1,d'],
\end{align*}
}}where the coefficients in $e_{p,q,n}(\tau)$ are $b_{1,t}=\frac{2^{-2t-1}}{t!\left(t+1\right)!}$,
$b_{2,t}=b_{1,t}\left[-\ln\left(2\right)-\frac{1}{2}\left(\psi\left(t+1\right)+\psi\left(t+2\right)\right)\right]$,
$t=n-1$, and $\psi\left(x\right)$ is the digamma function \cite{Table}.

\paragraph{$P_{out,CS}^{UB}$}

Then the high SNR approximation for $P_{out,CS}^{UB}\left(\rho_{th}\right)$
is given as {\small{{
\begin{equation}
\left\{ \begin{array}{cc}
G_{CS,1}^{UB}\left(\frac{2\rho_{th}}{\mathrm{SNR}}\right)^{d_{CS}^{LB}}\ln\left(\frac{\mathrm{SNR}}{2\rho_{th}}\right), & M_{\left[1\right]}=M_{\left[2\right]}=M_{\left[3\right]},\\
G_{CS,2}^{UB}\left(\frac{2\rho_{th}}{\mathrm{SNR}}\right)^{d_{CS}^{LB}}\ln\left(\frac{\mathrm{SNR}}{2\rho_{th}}\right), & M_{\left[1\right]}\geq M_{\left[2\right]}=M_{\left[3\right]},\\
G_{CS,3}^{UB}\left(\frac{2\rho_{th}}{\mathrm{SNR}}\right)^{d_{CS}^{LB}}, & M_{\left[1\right]}=M_{\left[2\right]}\geq M_{\left[3\right]},\\
G_{CS,4}^{UB}\left(\frac{2\rho_{th}}{\mathrm{SNR}}\right)^{d_{CS}^{LB}}, & M_{\left[1\right]}\neq M_{\left[2\right]}\neq M_{\left[3\right]},
\end{array}\right.\label{eq:UB_High}
\end{equation}
}}}where the diversity LB is $d_{CS}^{LB}=M_{\left[3\right]}$, and
the power gains are $G_{CS,1}^{UB}=6M_{[3]}3^{3M_{[3]}},$ $G_{CS,2}^{UB}=2M_{[3]}3^{3M_{[3]}}$,
$G_{CS,3}^{UB}=\varphi\left(M_{\left[2\right]},M_{\left[3\right]},3\right)+\varphi\left(M_{\left[3\right]},M_{\left[2\right]},3\right)$,
and $G_{CS,4}^{UB}=\sum_{l=1}^{2}\varphi\left(M_{\left[l\right]},M_{\left[3\right]},3\right)+\varphi\left(M_{\left[3\right]},M_{\left[l\right]},3\right)$.

\subsubsection{High SNR Approximations of \textmd{$P_{out,GS}^{LB}$ and $P_{out,GS}^{UB}$}}

\paragraph{$P_{out,GS}^{LB}$}

As shown in (\ref{eq:LB}), $P_{out,GS}^{LB}\left(\rho_{th}\right)=P_{out,CS}^{LB}\left(\rho_{th}\right)$,
when $M_{i\in[1,3]}=M$, then the high SNR approximation regarding
$P_{out,CS}^{LB}\left(\rho_{th}\right)$ in (\ref{eq:LB_High}) can
be modified to obtain
{\small\begin{equation}
P_{out,GS}^{LB}\left(\rho_{th}\right)\approx G_{GS}^{LB}\left(\frac{2\rho_{th}}{\mathrm{SNR}}\right)^{d_{GS}^{UB}}\ln\left(\frac{\mathrm{SNR}}{2\rho_{th}}\right),\label{eq:LB_High_G}
\end{equation}}where $d_{GS}^{UB}=M$ and $G_{GS}^{LB}=M3^{M}.$

\paragraph{$P_{out,GS}^{LB}$}

The derivations of the high SNR approximation of $P_{out,GS}^{UB}\left(\rho_{th}\right)$ are given
in Appendix V, and the result is given by
{\small\begin{equation}
P_{out,GS}^{UB}\left(\rho_{th}\right)\approx G_{GS}^{UB}\left(\frac{2\rho_{th}}{\mathrm{SNR}}\ln\left(\frac{\mathrm{SNR}}{2\rho_{th}}\right)\right)^{d_{GS}^{LB}},\label{eq:UB_High_G}
\end{equation}}where $d_{GS}^{UB}=M$ and $G_{GS}^{UB}=\left(6\cdot3^{3}\right)^{M}$.

%Based on the above analysis, the maximum MuD orders are obtained as $d_{CS}^{*}=\min(M_{1},M_{2},M_{3})$
%for both distributed and centralized CS, and $d_{GS}^{*}=M$ for both
%distributed and centralized GS. If the random scheduling is employed,
%the maximum MuD order would be just 1. Therefore, the proposed schemes
%obtain scalable MuD orders. Moreover, by showing that both the outage-optimal
%centralized scheduling and the proposed distributed scheduling achieve
%the same MuD orders, the optimality of the proposed distributed scheduling
%schemes are established with ER-UA transmission. Finally, given a
%total number of candidates, it is shown that the symmetric user configuration
%is most efficient to achieve a certain MuD order, which equally distributes
%the candidates in three clusters.

Based on the above analysis, the maximum MuD orders are obtained as $d_{CS}^{*}=\min(M_{1},M_{2},M_{3})$
for both distributed and centralized CS, and $d_{GS}^{*}=M$ for both
distributed and centralized GS. If the random scheduling is employed,
the maximum MuD order would be just 1. Therefore, the proposed schemes
obtain scalable MuD orders. 

%Moreover, by showing that both the outage-optimal
%centralized scheduling and the proposed distributed scheduling achieve
%the same DMT, the optimality of the proposed distributed scheduling
%schemes are established with ER-UA transmission. Finally, given a
%total number of candidates, it is shown that the symmetric user configuration
%is most efficient to achieve a certain MuD order, which equally distributes
%the candidates in three clusters.

\subsubsection{DMT Analysis}

In this subsection, the results of high SNR approximations are further
generalized and unified within the DMT framework \cite{zheng2003diversity}.
The DMT analysis gives a comprehensive description of the tradeoff
between transmission reliability and spectral efficiency with adaptive
data-rate. For the whole system with adaptive data-rate, a target
sum-rate is defined as ${\color{black}R_{th}}\left(\mathrm{SNR}\right)=r\log_{2}\left(1+\mathrm{SNR}\right)$
and $r$ is the multiplexing gain. Then, the outage probability of
the system is redefined with $R_{th}$ as
{\small\begin{equation}
P_{out}\left(\mathrm{SNR}\right)=\Pr\left(R\left(\mathrm{SNR}\right)\leq{\color{black}R_{th}}\left(\mathrm{SNR}\right)\right),\label{eq:Out_Rate}
\end{equation}}where $R\left(\mathrm{SNR}\right)=\frac{1}{2}\sum_{k\in[1,3]}\sum_{l\in\mathcal{L}_{k}}\log_{2}\left(1+\rho_{k,l}\left(\mathrm{SNR}\right)\right)$
is the instantaneous sum-rate of the system, and $\rho_{k,l}$ is
the end-to-end post-processing SNR of link $\mathsf{\mathsf{S}}_{l}\rightarrow\mathsf{R}\rightarrow\mathsf{\mathsf{S}}_{k}$.
Finally, the DMT is defined as
{\small\begin{equation}
d\left(r\right)=-\lim_{\mathrm{SNR}\rightarrow\infty}\frac{\log\left(P_{out}\left(\mathrm{SNR}\right)\right)}{\log\left(\mathrm{SNR}\right)}.\label{eq:DMT}
\end{equation}}After these definitions, a proposition is given to summarize the DMT
results.
\begin{prop}
\emph{Using ER-UA transmission, both centralized CS and distributed
CS achieve the same DMT as
{\small\begin{equation}
d_{CS}\left(r\right)=\min\left(M_{1},M_{2},M_{3}\right)\left(1-r/3\right)^{+},\label{eq:DMT1}
\end{equation}}and both centralized GS and distributed GS achieve the same DMT as
{\small\begin{equation}
d_{GS}\left(r\right)=M\left(1-r/3\right)^{+}.\label{eq:DMT2}\end{equation}}
}\end{prop}
\begin{IEEEproof}
See Appendix VI.
\end{IEEEproof}

\emph{Remark} 10: The maximum MuD orders are obtained as $d_{CS}^{*}=\min(M_{1},M_{2},M_{3})$
for both distributed and centralized CS, and $d_{GS}^{*}=M$ for both
distributed and centralized GS. If the random scheduling is employed,
the maximum MuD order would be just 1. Therefore, the proposed schemes
obtain scalable MuD orders. Moreover, by showing that both the outage-optimal
centralized scheduling and the proposed distributed scheduling achieve
the same DMT, the optimality of the proposed distributed scheduling
schemes are established with ER-UA transmission. Finally, given a
total number of candidates, it is shown that the symmetric user configuration
is most efficient from the DMT's perspective, which equally distributes
the candidates in three clusters.

\section{Numerical Results}

In this section, numerical results are presented to show the effectiveness
of the proposed schemes and validate the theoretical derivations.
The i.i.d. Rayleigh fading channels are assumed. The overall outage
probability is used as an effective metric to evaluate the transmission
reliability of the network. Specifically, the SNR threshold is set
as $\rho_{th}=1$ for each unicast stream, which corresponds to a
target rate of \textcolor{black}{$0.5\times6N\times\log_{2}\left(1+\rho_{th}\right)=3N$}
bit per channel use of the MIMO-Y channel. The symmetric SNR is assumed
for the relay system as $P_{T}/\sigma_{R}^{2}=P_{R}/\sigma_{S}^{2}=\mathrm{SNR}$.
The triplet $\left(M_{1},M_{2},M_{3}\right)$ is used to represent
the number of users in all the three clusters, and it is simplified
as $(M)$ for the GS.

\subsection{Performance and Complexity Comparisons}

\textcolor{black}{In this test case, we first compare both the distributed
and the centralized scheduling schemes for the Min-UA and the ER-UA
MIMO-Y transmissions with $N=1$. Then we present the performance
of distributed scheduling schemes for both transmissions with $N=2$ to demonstrate the effectiveness of the
proposed schemes in more general MIMO Y channels.} The symmetric user
configuration, i.e., $\left(M,M,M\right)$, is assumed. The random
selection is used as a reference, which achieves a MuD order of 1
regardless the user configuration. Therefore, the random selection
is also indicated by user configuration (1,1,1) for CS or simplified
as (1) for GS in the related figures.

\subsubsection{Min-UA Transmission \textcolor{black}{for $N=1$}}

%%%%%%%%%%%%%%%%%%%%%%%%%%%%%%%%%%%
\begin{figure}[t]
\centering \includegraphics[width=9cm]{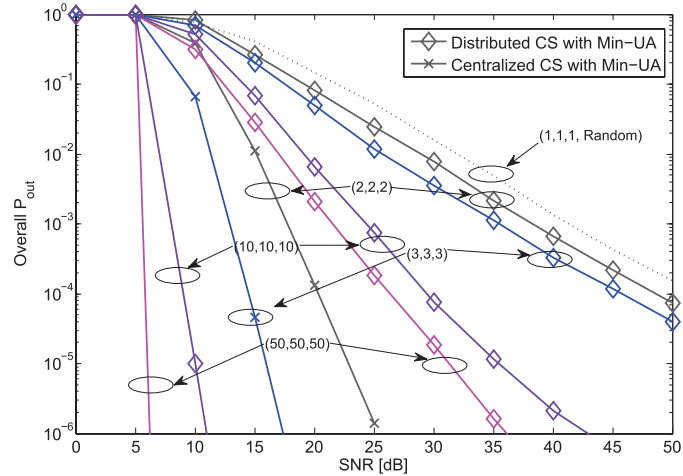} \protect\caption{Overall outage probability of the centralized and distributed CS with
Min-UA transmission, where $N_{R}=3,\, N_{T}=2$ and $\rho_{th}=1$.}
\end{figure}

%%%%%%%%%%%%%%%%%%%%%%%%%%%%%%%%%%%

\begin{figure}[t]
\centering \includegraphics[width=9cm]{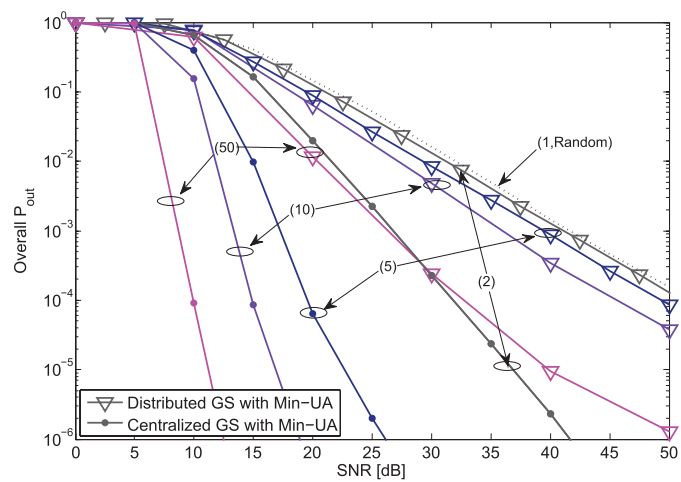} \protect\caption{Overall outage probability of the centralized and distributed GS with
Min-UA transmission, where $N_{R}=3,\, N_{T}=2$ and $\rho_{th}=1$.}
\end{figure}

Focusing on the Min-UA transmission, Fig. 6 and Fig. 7 present the
overall outage performances of the CS and the GS, respectively. It
is observed that the proposed distributed scheduling schemes are inferior
to the centralized schemes. Only when $M$ is relatively large, the
distributed scheduling shows distinctive performance improvement as
compared to the random selection.

%%%%%%%%%%%%%%%%%%%%%%%%%%%%%%%%%%%

\subsubsection{ER-UA Transmission \textcolor{black}{for $N=1$}}

Fig. 8 and Fig. 9 present the overall outage performances of CS and
GS with ER-UA transmission. As shown in these figures, the proposed
distributed user scheduling schemes and their centralized counterparts
achieve comparable performances. Moreover, a scalable MuD order is
observed for both the centralized scheduling and the proposed distributed
scheduling. These observations prove the optimality of the distributed
scheduling in terms of MuD order. %%%%%%%%%%%%%%%%%%%%%%%%%%%%%%%%%%%
\begin{figure}[t]
\centering \includegraphics[width=9cm]{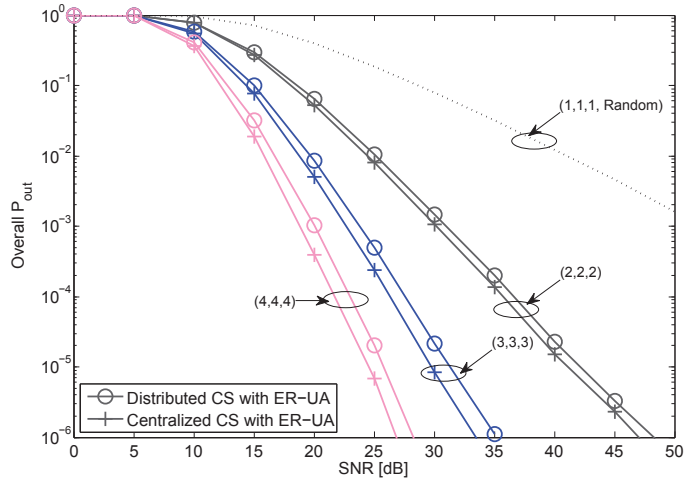} \protect\caption{Overall outage probability of the centralized and distributed CS with
ER-UA transmission, where $N_{R}=N_{T}=3$ and $\rho_{th}=1$.}
\end{figure}

%%%%%%%%%%%%%%%%%%%%%%%%%%%%%%%%%%%
\begin{figure}[t]
\centering \includegraphics[width=9cm]{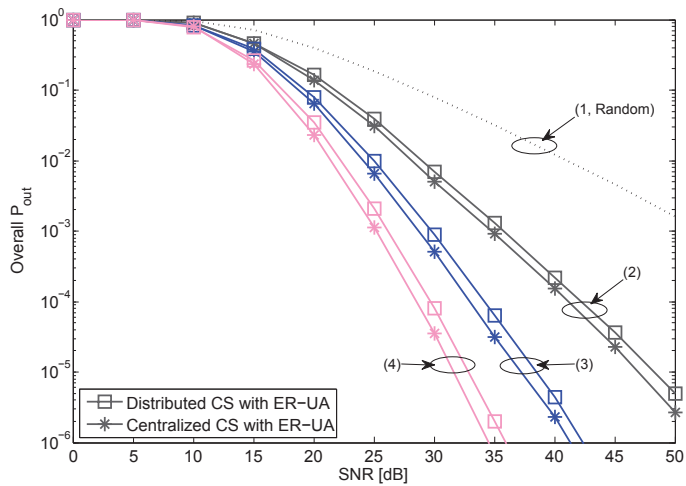} \protect\caption{Overall outage probability of the centralized and distributed GS with
ER-UA transmission, where $N_{R}=N_{T}=3$ and $\rho_{th}=1$.}
\end{figure}

%%%%%%%%%%%%%%%%%%%%%%

\subsubsection{\textcolor{black}{Min-UA and ER-UA Transmissions for $N=2$}}

\textcolor{black}{In order to validate the effectiveness of the proposed
distributed scheduling schemes in more general MIMO Y channels, Fig.
10 presents the overall outage performances of the distributed scheduling
schemes for Min-UA and ER-UA MIMO-Y transmissions with $N=2$. {\color{black}For
the Min-UA transmission, it is shown that the performance gains of
CS and GS are not significant. Especially, in the GS case the performance
improvement is limited.} On the contrary, significant performance
gain is achieved in CS and GS with ER-UA transmission, where a scalable
MuD order is observed.} %\begin{figure}[t]%\begin{centering}%\includegraphics[width=9cm]{Fig_11}%\par\end{centering}%\caption{\textcolor{black}{Overall outage probability of the distributed scheduling%schemes for the Min-UA and the ER-UA transmissions, where $N=2$ and%$\rho_{th}=1$.}}%\end{figure}

%%%%%%%%%%%%%%%%%%%%%%%%%%%%%%%%%%%
\begin{figure}[t]
\centering \includegraphics[width=9cm]{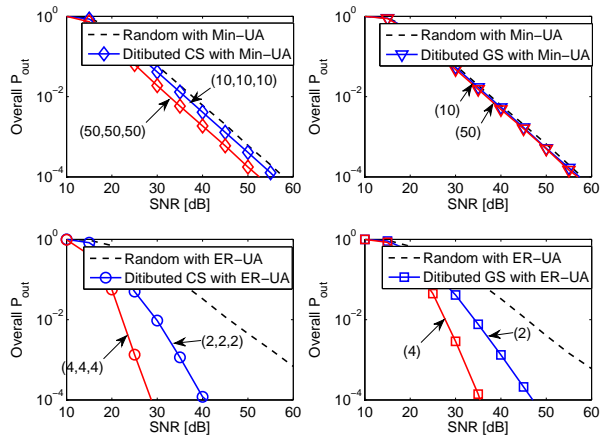} \protect\caption{Overall outage probability of the distributed scheduling schemes for
the Min-UA and the ER-UA transmissions, where $N=2$ and $\rho_{th}=1$.
Note that we set $N_{R}=3N=6,N_{T}=2N=4$ for Min-UA and $N_{R}=N_{T}=3N=6$
for ER-UA.}
\end{figure}

%%%%%%%%%%%%%%%%%%%%%%

\subsubsection{Complexity and CSI Overhead}

To further appreciate the proposed distributed scheduling, the computational
complexities and the CSI overheads of all the considered scheduling
schemes \textcolor{black}{are briefly analyzed} and compared. The complexity is measured in
terms of the number of floating point operations (flops) \cite{golub2012matrix}. In particular, the complexity is presented in a manner such that the distributed nature of our proposed scheme can be highlighted. {\color{black} It is noted, by using the big O notation, we can subsume the computation with respect to the SSA, matrix inversion and etc. We only maintain the key parameter $M$ and some necessary constants ($a_1$, $a_2$, $b_1$ and $b_2$) regarding the calculations of user/group scheduling metrics to highlight key parameters. The detailed analysis is collected in Appendix VII.}
As shown in Table I, the centralized scheduling schemes require global
CSI of all the candidates, which involves high CSI feedback overheads.
The computational complexity at the scheduling center is also relatively
high, as shown in Table II. In contrast to the centralized scheduling,
the proposed distributed methods allow each user to calculate its
own scheduling metric. For GS, such metric is explicitly fed back;
for CS, only the orders of these metrics are relevant and no explicitly
feedback of of CSI is necessary. Therefore, the distributed schemes significantly
reduce the computational complexity at the relay. It is also noted
that Min-UA and ER-UA transmission schemes show different performance-complexity
tradeoffs. For the Min-UA transmission, the low implementation complexity
is achieved at the cost of insufficiently utilized MuD gain; only
when the number of users is large, the distributed scheduling shows
distinctive performance improvement. On the other hand, the ER-UA
transmission allows the near-optimal distributed scheduling; therefore,
it is sufficient to apply the distributed scheduling to effectively
harvest the full MuD gains.

%%%%%%%%%%%%%%%%%%%%%%%%%%%%%%%%%%%
{
\begin{table*}[tp]
\tabcolsep 1.0mm \caption{CSI Overhead Analysis for Different Scheduling Schemes}
\label{Table2} \centering
{\scriptsize{
\begin{tabular}{|c|c|c|c|c|c|}
\hline
\multirow{2}{*}{\scriptsize \textbf{\textit{Schemes}}} & \multirow{2}{*}{\scriptsize \textbf{\textit{Node}}} & \multicolumn{2}{c|}{\scriptsize \textbf{\textit{CS}}} & \multicolumn{2}{c|}{\scriptsize \textbf{\textit{GS}}}\\
\cline{3-6}
 &  & {\scriptsize{{\textbf{\textit{Centralized}} }}}  & {\scriptsize{{\textbf{\textit{Distributed}} }}}  & {\scriptsize{{\textbf{\textit{Centralized}} }}}  & {\scriptsize{{\textbf{\textit{Distributed}}}}}\\
\hline
\multirow{2}{*}{\scriptsize{{\textbf{\textit{Min-UA}}}}}  & {\scriptsize{{\textbf{\textit{User}} }}}  & {\scriptsize{No CSI}}  & {\scriptsize{{Local CSI}}}  & {\scriptsize{{No CSI}}}  & {\scriptsize{{Local CSI}}}\\
\cline{2-6}
 & {\scriptsize{{\textbf{\textit{Relay}} }}}  & {\scriptsize{{Global CSI}}}  & {\scriptsize{{No CSI }}}  & {\scriptsize{{Global CSI}}}  & {\scriptsize{{\color{black}2bits + 1 feedback (with a probably of 2/9) of the angle in (17)}}}\\
\hline
\multirow{2}{*}{\scriptsize{{\textbf{\textit{ER-UA}}}}}  & {\scriptsize{{\textbf{\textit{User}} }}}  & {\scriptsize{{No CSI }}}  & {\scriptsize{{Local CSI}}}  & {\scriptsize{{No CSI}}}  & {\scriptsize{{Local CSI}}}\tabularnewline
\cline{2-6}
 & {\scriptsize{{\textbf{\textit{Relay}} }}}  & {\scriptsize{{Global CSI}}}  & {\scriptsize{{No CSI }}}  & {\scriptsize{{Global CSI}}}  & {\scriptsize{{\color{black}1 feedback of the angle in (17)}}}\tabularnewline
\hline
\end{tabular}}}
\end{table*}

\begin{table*}[tp]
\tabcolsep 1.0mm \caption{Complexity Analysis for Different Scheduling Schemes}
\label{Table1} \centering
{\scriptsize{
\begin{tabular}{|c|c|c|c|c|c|}
\hline
\multirow{2}{*}{\scriptsize \textbf{\textit{Schemes}}} & \multirow{2}{*}{\scriptsize \textbf{\textit{Node}}} & \multicolumn{2}{c|}{\scriptsize \textbf{\textit{CS}}} & \multicolumn{2}{c|}{\scriptsize \textit{GS}}\\
\cline{3-6}
 &  & {\scriptsize{{\textbf{\textit{Centralized}}}}}  & {\scriptsize{{\textbf{\textit{Distributed}}}}}  & {\scriptsize{{\textbf{\textit{Centralized}}}}}  & {\scriptsize{{\textbf{\textit{Distributed}}}}}\\
\hline
\multirow{2}{*}{\scriptsize{{\textbf{\textit{Min-UA}}}}}  & {\scriptsize{{\textbf{\textit{User}}}}}  & {\scriptsize{{-}}}  & {\scriptsize{{$O\left( {{1}} \right)$}}}  & {\scriptsize{{-}}}  & {\scriptsize{{$O\left( {{1}} \right)$}}}\\
\cline{2-6}
 & {\scriptsize{{\textbf{\textit{Relay}} }}}  & {\scriptsize{${O\left( {{M^3}} \right)}$}}  & {\scriptsize{{-}}}  & {\scriptsize{{$O\left( {{a_1}M} \right)$}}}  & {\scriptsize{{$O\left( {{a_2}M} \right)$}}}\\
\hline
\multirow{2}{*}{\scriptsize{{\textbf{\textit{ER-UA}}}}}  & {\scriptsize{{\textbf{\textit{User}}}}}  & {\scriptsize{{-}}}  & {\scriptsize{{$O\left( {{1}} \right)$}}}  & {\scriptsize{{-}}}  & {\scriptsize{{$O\left( {{1}} \right)$}}}\tabularnewline
\cline{2-6}
 & {\scriptsize{{\textbf{\textit{Relay}}}}}  & {\scriptsize{{${O\left( {{M^3}} \right)}$}}}  & {\scriptsize{{-}}}  & {\scriptsize{{$O\left( {{b_1}M} \right)$}}}  & {\scriptsize{{$O\left( {{b_2}M} \right)$}}}\tabularnewline
\hline
\end{tabular}

}\textcolor{black}{It is noted that ${a_1} \gg {a_2}$ and ${b_1} \gg {b_2}$, since all computation is done by the relay with centralized scheduling. "-" means no computation.}
}
\end{table*}

}

\subsection{Validating Theoretical Derivations for ER-UA Transmission}

Fig. 11 and Fig. 12 validate the theoretical derivations of the outage
probability bounds and the corresponding high SNR approximations for
CS and GS with $N=1$, respectively. It is shown that the derived bounds in Proposition
1, 2, and the relations in (\ref{eq:CC_Bound}) and (\ref{eq:GC_Bound})
are correct. It is also noted that the developed bounds are loose
due to several approximations, e.g., using the minimum eigenvalue
to obtain the upper bounds and relaxing the interval of integration
for more tractable results and etc. Fortunately, these bounds are still
useful and correct, because they enable the tractable and explicit
MuD order in Proposition 3. As verified in the figures, both asymptotic
results regarding the UB and LB show the same diversity order, accurately
bounding the MuD orders of both the centralized and distributed scheduling
schemes with CS and GS, respectively.
\begin{figure}[t]
\begin{centering}
\includegraphics[width=9cm]{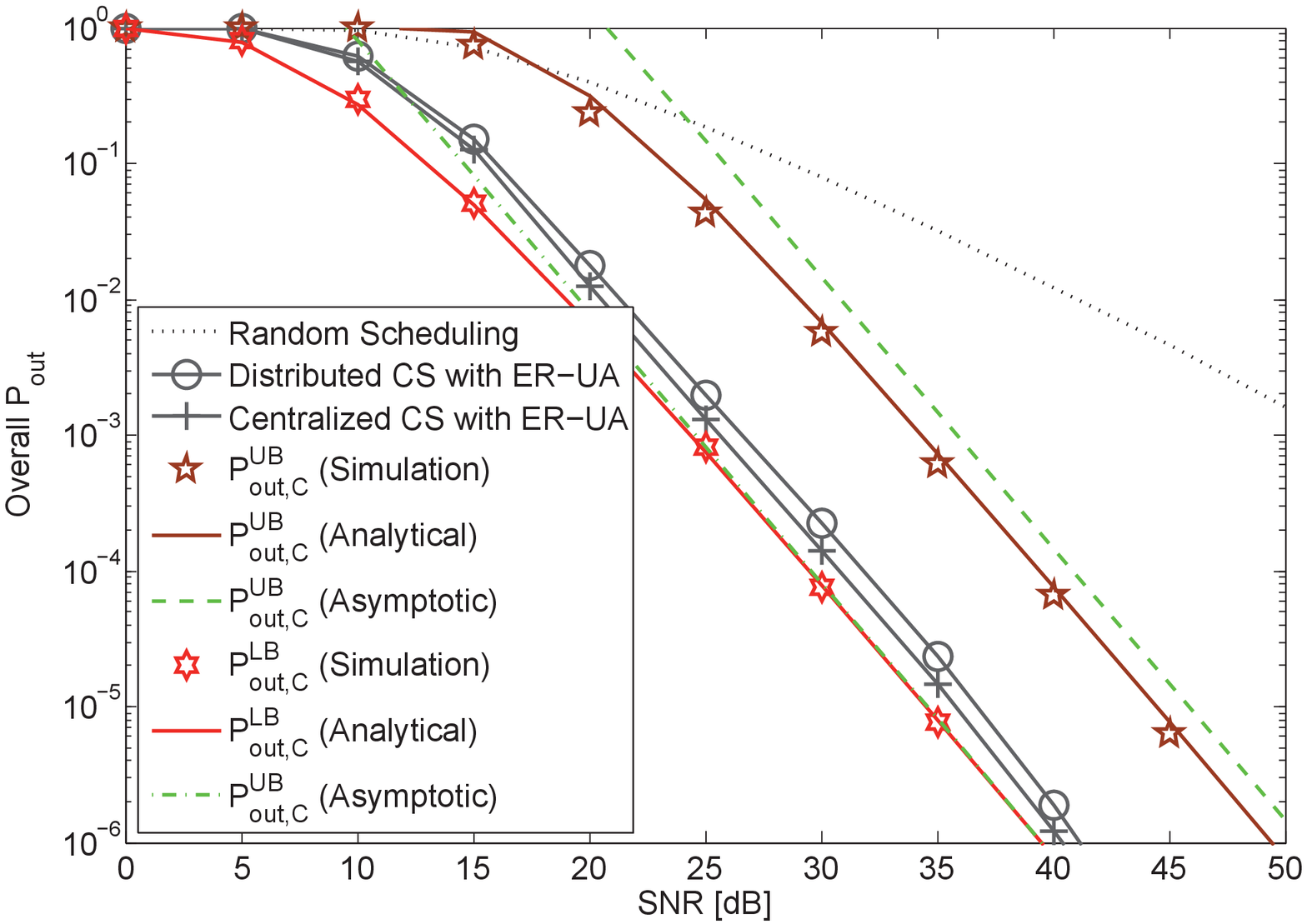}
\par\end{centering}

\caption{Overall outage probabilities of the centralized/distributed CS and
their bounds with high SNR approximations, where the user configuration
is $\left(2,3,4\right)$ and SNR threshold is $\rho_{th}=1$.}
\end{figure}

\begin{figure}[t]
\begin{centering}
\includegraphics[width=9cm]{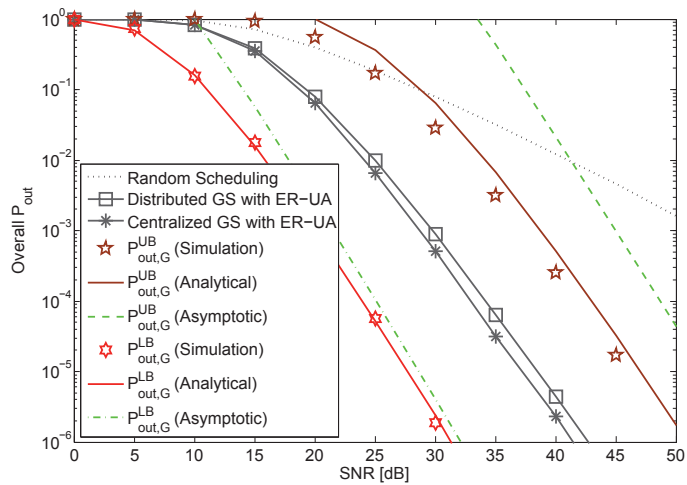}
\par\end{centering}

\caption{Overall outage probabilities of the centralized/distributed GS and
their bounds with high SNR approximations, where the user group configuration
is $M=3$ and the SNR threshold is $\rho_{th}=1$. }
\end{figure}

\section{Conclusion}

The distributed cluster-wise scheduling (CS) and group-wise scheduling
(GS) have been studied for the MIMO-Y channel with two transmission
schemes which have different requirements on the minimum number of
user antennas. The RSS has been employed to guide the distributed
CS and GS with the Min-UA transmission; and these low-complexity distributed
scheduling schemes obtain notable MuD gains when the candidates are
abundant. With a simpler yet effective implementation, the RSS-based
ER-UA MIMO-Y transmission has been proposed, and the corresponding
distributed CS and GS are theoretically proved to achieve the comparable
performances as their centralized counterparts. By comparing a variety
of scheduling schemes with Min-UA and ER-UA transmissions, the performance-complexity
tradeoffs of user scheduling has been revealed for the MIMO-Y channel.
Moreover, analysis with ER-UA transmission shows that the achievable
MuD gain is limited by the minimum number of users in the three clusters,
which sheds light into the fundamental behavior of MuD in the MIMO-Y
channel. Extending the distributed scheduling to the more general
{\color{black}multi-way relay channels} is a promising future work,
while the analysis for the explicit MuD behaviors with the Min-UA
transmission is still open. {\color{black}Moreover, it is noted that
the proposed scheme is based on a simple system model. Recently, some
new MIMO channel modeling methods are reported in \cite{cheng2009adaptive,cheng2011new}.
Studying the distributed user scheduling and the optimal beamforming
for these new channel models has more practical value and will be
of interest for future research.}

\appendix

\section*{Appendix I Lemmas}
{\color{black}\begin{lem}
\emph{The equivalent channel coefficient $\alpha_{m,j_{k}}=\left\Vert \mathbf{H}_{j_{k}}^{-1}\mathbf{e}_{m}\right\Vert ^{-1}$
with $\forall j_{}\in[1,M_{k}]$, $k\in[1,3]$ and $\forall m\in\left\{ \pi\left(l,k\right):\, l\text{\ensuremath{\in}}\mathcal{L}_{k}\right\} $
is lower bounded by the minimum eigenvalue of $\mathbf{H}_{j_{k}}\mathbf{H}_{j_{k}}^{H}$,
i.e., $\tilde{\lambda}_{j_{k}}:=\lambda_{min}\left(\mathbf{H}_{j_{k}}\mathbf{H}_{j_{k}}^{H}\right)\leq\alpha_{m,j_{k}}$\label{lem:ECG}}\end{lem}
\begin{IEEEproof}
Note that $\left\Vert \mathbf{e}_{m}\right\Vert =1$, then according
to Rayleigh-Ritz theorem \cite{golub2012matrix} the lemma can be proved. \end{IEEEproof}

\begin{lem}
\emph{For two objective functions $\eta_{A}$ and $\eta_{B}$ with respect
to the selected users $J=\left\{ j_{1},j_{2},j_{3}\right\} $, the
optimal selections are defined as $J_{A}=\arg\thinspace\max_{J\in\mathcal{J}}\thinspace\left(\eta_{A}^{\left(J\right)}\right)$
and $J_{B}=\arg\thinspace\max_{J\in\mathcal{J}}\thinspace\left(\eta_{B}^{\left(J\right)}\right)$
respectively, then $\eta_{A}^{\left(J_{B}\right)}\leq\eta_{A}^{\left(J_{A}\right)}$
and $\eta_{B}^{\left(J_{A}\right)}\leq\eta_{B}^{\left(J_{B}\right)}$.
\label{lem:OBJ}}\end{lem}
\begin{IEEEproof}
Based on its definition, $J_{A}$ maximizes $\eta_{A}^{\left(J\right)}$
over $J\in\mathcal{J}$, therefore no any other $J'\in\mathcal{J}$,
$J'\neq J_{A}$ can achieve larger $\eta_{A}^{\left(J'\right)}$ than
$\eta_{A}^{\left(J_{A}\right)}$. Let $J'=J_{B}$, $\eta_{A}^{\left(J_{B}\right)}\leq\eta_{A}^{\left(J_{A}\right)}$
is proved. {\color{black}For the same reason $\eta_{B}^{\left(J_{A}\right)}\leq\eta_{B}^{\left(J_{B}\right)}$}
can be proved.\end{IEEEproof}
\begin{lem}
\emph{For a set $\left\{ x_{1},x_{2},x_{3}\right\} $, $x_{i}>0$, $i\in[1,3]$
define an ordered set $\mathcal{X}$ containing all 2-element permutation
of $\left\{ x_{1},x_{2},x_{3}\right\} $, i.e., $\mathcal{X}=\left\{ \left(x_{1},x_{2}\right)\thinspace\left(x_{2},x_{1}\right)\thinspace...\thinspace\left(x_{2},x_{3}\right)\thinspace\left(x_{3},x_{2}\right)\right\} $,
and let $x_{[n]}$ as the $n$-th largest element of $\left\{ x_{1},x_{2},x_{3}\right\} $.
Then the following inequality holds
\[
g\left(x_{\left[3\right]},x_{\left[2\right]}\right)\leq\min_{\left(\tilde{x}_{1},\tilde{x_{2}}\right)\in\mathcal{X}}\left(g\left(\tilde{x}_{1},\tilde{x_{2}}\right)\right),
\]
where $g\left(\tilde{x}_{1},\tilde{x_{2}}\right):=\frac{\tilde{x}_{1}\tilde{x_{2}}\mathrm{SNR}}{2\tilde{x_{2}}+6\left(1+\mathrm{SNR}^{-1}\right)}$
is defined over $\mathcal{X}$ and monotonically increases with $\tilde{x}_{1}$
and $\tilde{x}_{2}$. \label{lem:min}}\end{lem}
\begin{IEEEproof}
Note that $g\left(\tilde{x}_{1},\tilde{x_{2}}\right)$ monotonically
increases with $\tilde{x}_{1}$ and $\tilde{x}_{2}$, we can check that $\min_{\left(\tilde{x}_{1},\tilde{x_{2}}\right)\in\mathcal{X}}\left(g\left(\tilde{x}_{1},\tilde{x_{2}}\right)\right)=\min\left(g\left(x_{\left[3\right]},x_{\left[2\right]}\right),g\left(x_{\left[2\right]},x_{\left[3\right]}\right)\right)=g\left(x_{\left[3\right]},x_{\left[2\right]}\right),$
and the proof is finished.
\end{IEEEproof}

\begin{lem} \emph{Given $\mathbf{e}_{m}^{[n]}$, a reference direction
from the orthogonal basis $\mathbf{E}$ of the RSS, and the channel
matrix $\mathbf{H}_{j_{k}}$ with i.i.d. $\mathcal{CN}\left(0,\,1\right)$
entries, the equivalent channel gain $\left(\alpha_{m,j_{k}}^{[n]}\right)^{2}=\left\Vert \mathbf{H}_{j_{k}}^{-1}\mathbf{e}_{m}^{[n]}\right\Vert ^{-2}$
of the $n$-th data stream within the link $\mathsf{\mathsf{S}}_{j_{k}}\rightarrow\mathsf{R}$
is exponentially distributed as $f_{\left(\alpha_{m,j_{k}}^{[n]}\right)^{2}}\left(x\right)=e^{-x}$.
}\end{lem} \begin{IEEEproof} Firstly, it is noted that $\mathbf{e}_{m}^{[n]}=\mathbf{E}\mathbf{a}_{m'}$,
where $\mathbf{a}_{m'}\in\mathbb{C}^{3N\times1}$ has all zero elements
except the $m'$-th element that equals to one. It is noted that $m'=\left(m-1\right)N+n$,
$m\in[1,3]$ and $n\in[1,N]$. Then $\left(\alpha_{m,j_{k}}^{[n]}\right)^{2}$
is re-expressed as {\small{}
\begin{eqnarray}
\left(\alpha_{m,j_{k}}^{[n]}\right)^{2} & = & \left\Vert \mathbf{H}_{k}^{-1}\mathbf{E}\mathbf{a}_{m'}\right\Vert ^{-2}=\left[\left(\mathbf{\tilde{H}}_{j_{k}}\tilde{\mathbf{H}}_{j_{k}}^{H}\right)^{-1}\right]_{m',m'}^{-1},\label{eq:ecg}
\end{eqnarray}
}Note that $\mathbf{H}_{j_{k}}\mathbf{H}_{j_{k}}^{H}$ and $\mathbf{\tilde{H}}_{j_{k}}\tilde{\mathbf{H}}_{j_{k}}^{H}=\mathbf{E}\mathbf{H}_{j_{k}}\mathbf{H}_{j_{k}}^{H}\mathbf{E}^{H}$
have the same statistic properties, because of the independence between
$\mathbf{E}$ and $\mathbf{H}_{j_{k}}$, and the unitarily invariant
of the central Wishart matrix \cite{tulino2004random}. According
to \cite{GoreISIT02}, we know $\left[\left(\mathbf{H}_{j_{k}}\mathbf{H}_{j_{k}}^{H}\right)^{-1}\right]_{m',m'}^{-1}$
is exponentially distributed with a PDF $f(x)=e^{-x}$, so is $\left(\alpha_{m,j_{k}}^{[n]}\right)^{2}$
as defined in (\ref{eq:ecg}), and the proof is finished. \end{IEEEproof}

\begin{lem}
$\left(\frac{c}{\mathrm{SNR}}\right)^{d}\ln\left(\frac{\mathrm{SNR}}{c}\right)\dot{=}\mathrm{SNR}^{-d}$\end{lem}
\begin{IEEEproof}
Using the l'Hospital's rule, it is easy to check that $\left(\frac{c}{\mathrm{SNR}}\right)^{d}$
$=o\left(\left(\frac{c}{\mathrm{SNR}}\right)^{d}\ln\left(\frac{\mathrm{SNR}}{c}\right)\right)$
and $\left(\frac{c}{\mathrm{SNR}}\right)^{d}\ln\left(\frac{\mathrm{SNR}}{c}\right)=o\left(\left(\frac{c}{\mathrm{SNR}}\right)^{d-\epsilon}\right)$
when $\mathrm{SNR}\rightarrow\infty$ with $\forall\epsilon>0$, and
the inequality
{\small\begin{align}
c_{1}\lim_{\mathrm{SNR}\rightarrow\infty}\left(\frac{c}{\mathrm{SNR}}\right)^{d} & <\lim_{\mathrm{SNR}\rightarrow\infty}\left(\frac{c}{\mathrm{SNR}}\right)^{d}\ln\left(\frac{\mathrm{SNR}}{c}\right)\nonumber \\
 & <c_{2}\lim_{\mathrm{SNR}\rightarrow\infty}\left(\frac{c}{\mathrm{SNR}}\right)^{d-\epsilon},\label{eq:ieq}
\end{align}}holds for some $c_{1}$ and $c_{2}$. Let $\epsilon\rightarrow0$
in (\ref{eq:ieq}), then $\left(\frac{c}{\mathrm{SNR}}\right)^{d-\epsilon}\rightarrow\left(\frac{c}{\mathrm{SNR}}\right)^{d}$
and $\left(\frac{c}{\mathrm{SNR}}\right)^{d}\ln\left(\frac{\mathrm{SNR}}{c}\right)\dot{=}\mathrm{SNR}^{-d}$,
and the proof is finished.\end{IEEEproof}

{\color{black}\section*{Appendix II Proof of Proposition 1}
\begin{IEEEproof}
For inequality $\left(a1\right)$, use Lemma \ref{lem:ECG} in Appendix I to obtain
$\tilde{\lambda}_{j_{k}}:=\lambda_{min}\left(\mathbf{H}_{j_{k}}\mathbf{H}_{j_{k}}^{H}\right)\leq\alpha_{m,j_{k}}^{2}$,
$\forall m\in\left\{ \pi\left(l,k\right):\, l\text{\ensuremath{\in}}\mathcal{L}_{k}\right\} $.
Therefore, $\tilde{\lambda}_{j_{k}}\leq\alpha_{j_{k}}^{2}:=\min_{m}\left\{ \alpha_{m,j_{k}}^{2}\right\} $
for $\forall j_{}\in\left[1,M_{k}\right]$, $k\in[1,3]$. By choosing $j_{k}=j_{k}^{\star}$,
we have
{\small\begin{equation}
\tilde{\lambda}_{j_{k}^{\star}}\leq\alpha_{j_{k}^{\star}}^{2}.\label{eq:par1}
\end{equation}}On the other hand, note that $j_{k}^{\ddagger}=\arg\thinspace\max_{j_{}\in[1,M_{k}]}\thinspace\left(\alpha_{j_{k}}^{2}\right)$, $k\in[1,3]$, i.e., $j_{k}^{\ddagger}$ is optimal for $\alpha_{j_{k}}^{2}$; by
using Lemma \ref{lem:OBJ} in Appendix I, it is observed
{\small\begin{equation}
\alpha_{j_{k}^{\star}}^{2}\leq\alpha_{j_{k}^{\ddagger}}^{2},\thinspace k\in[1,3].\label{eq:par2}
\end{equation}}Combining (\ref{eq:par1}) and (\ref{eq:par2}) we have
{\small\begin{equation}
\tilde{\lambda}_{j_{k}^{\star}}\leq\alpha_{j_{k}^{\ddagger}}^{2},\thinspace k\in[1,3].\label{eq:par3}
\end{equation}}Based on $J_{\lambda}=\left\{ \tilde{\lambda}_{j_{1}^{\star}},\thinspace\tilde{\lambda}_{j_{2}^{\star}},\thinspace\tilde{\lambda}_{j_{3}^{\star}}\right\} $,
we introduce $\zeta_{min}^{\left(J_{\lambda}\right)}:=\min_{k\in[1,3],\, l\text{\ensuremath{\in}}\mathcal{L}_{k}}\left(g\left(\tilde{\lambda}_{j_{k}^{\star}},\tilde{\lambda}_{j_{l}^{\star}}\right)\right)$,
where $g\left(x,y\right):=\frac{xy\mathrm{SNR}}{2x+6\left(1+\mathrm{SNR}^{-1}\right)}$.
It is also noted that $\rho_{min,CS}^{\left(J_{D}\right)}=\min_{k\in[1,3],\, l\text{\ensuremath{\in}}\mathcal{L}_{k}}\left(g\left(\alpha_{j_{k}^{\ddagger}}^{2},\alpha_{,j_{l}^{\ddagger}}^{2}\right)\right)$.
According to (\ref{eq:par3}) and the monotonicity of $g\left(x,y\right)$,
it is easy to check that{\small\begin{equation}
\zeta_{min}^{\left(J_{\lambda}\right)}\leq\rho_{min,CS}^{\left(J_{D}\right)}.\label{eq:par4}
\end{equation}}Finally, note that $\tilde{\lambda}_{[n]}^{\left(J_{\lambda}\right)}$
is the $n$-th largest element of $J_{\lambda}$, then according to
Lemma \ref{lem:min} in Appendix I, we can see that{\small\begin{equation}
\rho_{min,CS}^{LB}:=g\left(\tilde{\lambda}_{[3]}^{\left(J_{\lambda}\right)},\tilde{\lambda}_{[2]}^{\left(J_{\lambda}\right)}\right)\leq\zeta_{min}^{\left(J_{\lambda}\right)}.\label{eq:par5}
\end{equation}}Combining (\ref{eq:par5}) and (\ref{eq:par4}), the inequality $\left(a1\right)$
is proved. Next, according to Lemma \ref{lem:OBJ} in Appendix I, the inequality
$\left(a2\right)$ is straightforward by noting that $J_{C}$ is optimal
for $\rho_{min,CS}^{\left(J\right)}$. For inequality $\left(a3\right)$,
it is observed that
{\small\begin{equation}
\rho_{min,CS}^{\left(J_{C}\right)}\leq\rho_{j_{2}^{*},j_{1}^{*}}=g\left(\alpha_{\mathrm{I},j_{1}^{*}}^{2},\alpha_{\mathrm{I},j_{2}^{*}}^{2}\right)\leq g\left(\alpha_{\mathrm{I},j_{1}^{\dagger}}^{2},\alpha_{\mathrm{I},j_{2}^{\dagger}}^{2}\right),\label{eq:inequality_chain-1}
\end{equation}}where $\left\{ j_{1}^{*},j_{2}^{*}\right\} \subset J_{C}$ and the
last inequality is based on the monotonicity of $g\left(x,y\right)$
and Lemma \ref{lem:OBJ} in Appendix I, i.e., $\alpha_{\mathrm{I},j_{k}^{*}}^{2}\leq\alpha_{\mathrm{I},j_{k}^{\dagger}}^{2}$
and $j_{k}^{\dagger}$ is defined to be optimal for $\alpha_{\mathrm{I},j_{k}}^{2}$.
Then the proof is finished.
\end{IEEEproof}}

\section*{Appendix III Proof of Proposition 2}
{\color{black}\begin{IEEEproof}
Recall that $\rho_{j_{l}^{'},j_{k}^{'}}=g\left(\alpha_{m,j_{l}^{;}}^{2},\alpha_{m,j_{k}^{'}}^{2}\right)$
and $\alpha_{j'_{k}}^{2}=\min_{m\in\left\{ \mathrm{I,II,III}\right\} }\left\{ \alpha_{m,j'_{k}}^{2}\right\} $,
then according to the monotonicity of $g\left(x,y\right)$, we can arrive at
{\small\begin{equation}
\rho_{j_{l}^{'},j_{k}^{'}}\geq g\left(\alpha_{m,j_{l}^{;}}^{2},\alpha_{m,j_{k}^{'}}^{2}\right)\geq g\left(\alpha_{j_{l}^{;}}^{2},\alpha_{j_{k}^{'}}^{2}\right).\label{eq:ppar1}
\end{equation}}Let us denote $\left(\alpha_{[n]}^{\left(J'\right)}\right)^{2}$ and
$\tilde{\lambda}_{\left[n\right]}^{\left(J'\right)}$ as the $n$-th
largest elements of $\left\{ \alpha_{j'_{1}}^{2},\alpha_{j'_{2}}^{2},\alpha_{j'_{3}}^{2}\right\} $
and $\left\{ \tilde{\lambda}_{j'_{1}},\tilde{\lambda}_{j'_{2}},\tilde{\lambda}_{j'_{3}}\right\} $,
where $\tilde{\lambda}_{j_{k}}:=\lambda_{min}\left(\mathbf{H}_{j_{k}}\mathbf{H}_{j_{k}}^{H}\right)$.
Then we can construct a lower bound of $\rho_{min,GS}^{\left(J'\right)}=\min_{k\in\left[1,3\right],l\in\mathcal{L}_{k}}\left(\rho_{j_{l}^{'},j_{k}^{'}}\right)$
as
{\small\begin{align}
\rho_{min,GS}^{\left(J'\right)} & \geq\min_{k\in\left[1,3\right],l\in\mathcal{L}_{k}}\left(g\left(\alpha_{j_{l}^{;}}^{2},\alpha_{j_{k}^{'}}^{2}\right)\right)\nonumber \\
 & \geq g\left(\left(\alpha_{[3]}^{\left(J'\right)}\right)^{2},\left(\alpha_{[2]}^{\left(J'\right)}\right)^{2}\right):=\gamma_{min}^{\left(J'\right)}\nonumber \\
 & \geq g\left(\tilde{\lambda}_{\left[3\right]}^{\left(J'\right)},\tilde{\lambda}_{\left[2\right]}^{\left(J'\right)}\right):=\zeta_{min}^{\left(J'\right)},\label{eq:approximation1}
\end{align}}where the first inequality is based on (\ref{eq:ppar1}), the second
inequality is based on Lemma \ref{lem:min} in Appendix I and the third inequality
is based on Lemma \ref{lem:ECG} in Appendix I and the monotonicity of $g\left(x,y\right)$.
According to (\ref{eq:approximation1}), We show that
{\small\begin{equation}
\rho_{min,GS}^{\left(J'\right)}\geq\gamma_{min}^{\left(J'\right)}\geq\zeta_{min}^{\left(J'\right)},\,\forall J'\in\mathcal{J}',\label{eq:basic relation}
\end{equation}}then we proceed to obtain the following inequalities
{\small\begin{equation}
\rho_{min,GS}^{\left(J'_{D}\right)}\overset{}{\geq}\gamma_{min}^{\left(J'_{D}\right)}\overset{}{\geq}\gamma_{min}^{\left(J_{\varsigma}\right)}\overset{}{\geq}\zeta_{min}^{\left(J_{\varsigma}\right)}:=\rho_{min,GS}^{LB},\label{eq:eq_chains_ag}
\end{equation}}where $J_{\varsigma}$ is defined as $J_{\varsigma}=\arg\,\max_{J'\in\mathcal{J}'}\left\{ \zeta_{min}^{\left(J'\right)}\right\} $.
i.e., $\zeta_{min}^{\left(J_{\varsigma}\right)}=\max_{J'\in\mathcal{J}'}\left\{ g\left(\tilde{\lambda}_{\left[3\right]}^{\left(J'\right)},\tilde{\lambda}_{\left[2\right]}^{\left(J'\right)}\right)\right\} =\rho_{min,GS}^{LB}$.
Here, the first and the third inequalities in (55) are based on the basic relations
in (\ref{eq:basic relation}), and the second inequality is based on
the optimality of $J'_{D}$ for $\left\{ \gamma_{min}^{\left(J'\right)}\right\} $
and Lemma 2 in Appendix I. It is noted that the inequalities in (\ref{eq:eq_chains_ag})
have been used to prove the inequality (b1) in (\ref{eq:the bound-1}).
Then, the inequality $\left(b2\right)$ is based on the fact that
$J'_{C}$ is optimal for $\rho_{min}^{\left(J'\right)}$ and Lemma
\ref{lem:OBJ} in Appendix I, where $J'\in\mathcal{J}'$. For the inequality $\left(b3\right)$,
it is noted that
{\small\[
\rho_{min,GS}^{\left(J'_{C}\right)}\leq\rho_{j_{2}^{'*},j_{1}^{'*}}=g\left(\alpha_{\mathrm{I},j_{1}^{'*}}^{2},\alpha_{\mathrm{I},j_{2}^{'*}}^{2}\right)\leq g\left(\alpha_{\mathrm{I},j_{1}^{\dagger}}^{2},\alpha_{\mathrm{I},j_{2}^{\dagger}}^{2}\right)
\]}where $\left\{ j_{1}^{'*},j_{2}^{'*}\right\} \subset J'_{C}$ and
the last inequality is based on the monotonicity of $g\left(x,y\right)$
and Lemma \ref{lem:OBJ} in Appendix I, i.e., $\alpha_{\mathrm{I},j_{k}^{'*}}^{2}\leq\alpha_{\mathrm{I},j_{k}^{\dagger}}^{2}$
and $j_{k}^{\dagger}$ is defined to be optimal for $\alpha_{\mathrm{I},j_{k}}^{2}$.
\end{IEEEproof}}

\section*{Appendix IV Derivations for \textmd{\normalsize {{{{$P_{out,CS(GS)}^{LB}$,
$P_{out,CS}^{UB}$ }}}}}{\normalsize {{{and }}}}\textmd{\normalsize {{{{$P_{out,GS}^{UB}$}}}}}}

As the preliminary, the relevant distributions of the key random variables
(RVs) are given. First, according to Lemma 4 in Appendix I, it is
known that $\mathrm{\alpha}_{m,{j_{k}}}^{2}$ is exponentially distributed
with the probability density function (PDF) as $f_{\mathrm{\alpha}_{m,j_{k}}^{2}}\left(x\right)=e^{-x}$,
therefore, the order statistic \cite{david1970order} is applied to
these i.i.d. RVs, and we obtain the PDF of $\alpha_{m,j_{k}^{\ddagger}}^{2}$
as $f_{\alpha_{m,j_{k}^{\ddagger}}^{2}}\left(x\right)=M_{k}\left(1-e^{-x}\right)^{M_{k}-1}e^{-x}$.
According to \cite{edelmaneigenvalues}, $\tilde{\lambda}_{j_{k}}=\lambda_{min}\left(\mathbf{H}_{j_{k}}\mathbf{H}_{j_{k}}^{H}\right)$
is also exponentially distributed as $f_{\mathrm{\alpha}_{m,{j_{k}}}^{2}}\left(x\right)=e^{-3x}$,
therefore, the PDF of $\tilde{\lambda}_{j_{k}^{\star}}$ is given
by $f_{\tilde{\lambda}_{j_{k}^{\star}}}\left(x\right)=M_{k}\left(1-e^{-3x}\right)^{M_{k}-1}e^{-3x}$.
Then, we continue the derivations for $P_{out,CS(GS)}^{LB}$, $P_{out,CS}^{UB}$
and $P_{out,GS}^{UB}$.

\paragraph*{\textmd{$P_{out,CS(GS)}^{LB}$}}
Let us first define $X=\alpha_{\mathrm{I},j_{1}^{\dagger}}^{2}$ and $Y=\alpha_{\mathrm{I},j_{2}^{\dagger}}^{2}$,
then we can express the LB of outage probability as {\small {{{{{$P_{out,CS}^{LB}\left(\rho_{th}\right)=\int_{0}^{\infty}f_{Y}\left(y\right)\left(\int_{0}^{a+\frac{b}{y}}f_{X}\left(x\right)dx\right)dy$}}}}}},
where $a\left(\rho_{th}\right)=\frac{2\rho_{th}}{\mathrm{SNR}}$ and
$b\left(\rho_{th}\right)=\frac{6\rho_{th}\left(\mathrm{SNR}+1\right)}{\mathrm{SNR^{2}}}.$
After some straightforward calculations, $P_{out,CS}^{LB}\left(\rho_{th}\right)$
is given in (\ref{eq:LB}).

\paragraph*{\textmd{$P_{out,CS}^{UB}$}}

With a slight abuse of notations, let us define $X=\tilde{\lambda}_{j_{1}^{\star}}$,
$Y=\tilde{\lambda}_{j_{2}^{\star}}$ and $Z=\tilde{\lambda}_{j_{3}^{\star}}$,
and introduce the ordered set ${\Theta}=\left\{ \Theta_{i}\right\} _{i=1}^{6}$
to collect all the permutations among $\left\{ X,Y,Z\right\} $, where
$\Theta_{i}=\left\{ \theta_{i,n}\right\} _{n=1}^{3},$ e.g., $\Theta_{1}=\left\{ \theta_{1,1},\theta_{1,2},\theta_{1,3}\right\} =\left\{ X,Y,Z\right\} $,
$\Theta_{2}=\left\{ \theta_{2,1},\theta_{2,2},\theta_{2,3}\right\} =\left\{ X,Z,Y\right\} $
and so on. We also associate $\Theta_{i}$ with an event as $O_{i}$,
which orders the elements of $\Theta_{i}$, e.g., $\Theta_{1}=\left\{ X,Y,Z\right\} \leftrightarrow O_{1}=\left\{ X\geq Y\geq Z\right\}$. Then we divide the whole integral region into six subregions to facilitate the calculation, where each subregion is associated with $O_{i}$, $i\in\left[{1,6}\right]$. Based on such division, $P_{out,CS}^{UB}\left(\rho_{th}\right)$ is re-expressed as
{\small\begin{equation}
P_{out}^{UB}\left(\rho_{th}\right)=\sum_{i=1}^{6}\Pr\left\{ E_{i}\left(\rho_{th}\right)\right\} ,\label{UB}
\end{equation}}where $E_{i}\left(\rho_{th}\right)=\left\{ g\left(\theta_{i},_{[3]}\theta_{i,[2]}\right)\leq\rho_{th},\, O_{i}\right\}$, $\theta_{i},_{[n]}$ is the $n$-th largest element of $\Theta_{i}$.
Without loss of generality, let us focus on the probability of $E_{1}\left(\rho_{th}\right)$,
given by $\Pr\left\{ E_{1}\left(\rho_{th}\right)\right\} =\sum_{t=1}^{3}\int\int\int_{D_{1,t}}f\left(x,y,z\right)dxdydz=\sum_{t=1}^{3}I_{1,t}$,
where $f\left(x,y,z\right)=f_{X}\left(x\right)f_{Y}\left(y\right)f_{Z}\left(z\right)$
due to independence, and the integral region of $I_{1,t}$
is $D_{1,t}=\left\{ \left(x,y,z\right)|\, x\in\mathcal{X}_{1,t},y\in\mathcal{Y}_{1,t},\, z\in\mathcal{Z}_{1,t}\right\} $,
i.e., $\int_{\mathcal{X}_{1,t}}\int_{\mathcal{Y}_{1,t}}\int_{\mathcal{Z}_{1,t}}$.
Then $I_{1,1}=\int_{0}^{\mu}\int_{0}^{x}\int_{0}^{y}f\left(x,y,z\right)dxdydz$
and $I_{1,2}=\int_{\mu}^{\infty}\int_{0}^{\mu}\int_{0}^{y}f\left(x,y,z\right)dxdydz$
are obtained after some calculations as
{\small\begin{align}
I_{1,1} & =\frac{M_{2}M_{3}\left(1-e^{-3\mu}\right)^{M_{\Sigma}}}{\left(M_{1}+M_{2}\right)M_{\Sigma}},\label{eq:I1}
\end{align}}
{\small\begin{align}
I_{1,2} & =\frac{M_{2}\left(1-e^{-3\mu}\right)^{M_{1}+M_{2}}\left[1-\left(1-e^{-3\mu}\right)^{M_{3}}\right]}{M_{1}+M_{2}},\label{eq:I2}
\end{align}}where $\mu=\frac{1}{2}\left(a+\sqrt{a^{2}+4b}\right)$ is the positive
root of the quadratic equation $y^{2}=ay+b$, $a\left(\rho_{th}\right)=\frac{2\rho_{th}}{\mathrm{SNR}}$, $b\left(\rho_{th}\right)=\frac{6\rho_{th}\left(\mathrm{SNR}+1\right)}{\mathrm{SNR^{2}}}$ and $M_{\Sigma}=M_{1}+M_{2}+M_{3}$. It is note that $I_{1,3}=\int_{\mu}^{\infty}\int_{\mu}^{x}\int_{0}^{a+\frac{b}{y}}f\left(x,y,z\right)dxdydz$
is not easy to calculate with tractable results, thus the following
approximations is used as $I_{1,3}<I_{1,3}^{'}=\int_{\mu}^{\infty}\int_{\mu}^{\infty}\int_{0}^{a+\frac{b}{y}}f\left(x,y,z\right)dxdydz$,
where the integral region regarding variable $y$ has been enlarged.
After some calculations, the result of $I_{1,3}^{'}$ is given by
{\small\begin{equation}
I'_{1,3}=\left[3M_{2}\sum_{p=0}^{M_{2}-1}\sum_{q=0}^{M_{1}}\left(-1\right)^{q}C_{p}^{M_{2}-1}C_{q}^{M_{1}}e^{-3qa}\int_{\mu}^{\infty}e^{-3(\left(p+1\right)y-qb\frac{1}{y})}dy\right]\left[1-\left(1-e^{-3\mu}\right)^{M_{3}}\right],\label{eq:App1}
\end{equation}}then $\Pr\left\{ E_{1}\left(\rho_{th}\right)\right\} \approx I_{1,1}+I_{1,2}+I'_{1,3}$
is obtained. It is also noted that the permutation of $\left\{ X,Y,Z\right\}$ and the user configuration $\left\{ M_{1},M_{3},M_{3}\right\}$ have a fixed matching pattern in $E_{i}\left(\rho_{th}\right)$; therefore, we can modify the
results of $\Pr\left\{ E_{1}\left(\rho_{th}\right)\right\} $ in (\ref{eq:I1}),
(\ref{eq:I2}) and (\ref{eq:App1}) to get $\Pr\left\{ E_{i}\left(\rho_{th}\right)\right\}$,
$i\in[2,6]$, and finally arrive at the result in (\ref{eq:UB}).

\paragraph*{\textmd{$P_{out,GS}^{UB}$}}

Applying the order statistics of i.i.d. RVs, $P_{out,GS}^{UB}\left(\rho_{th}\right)$
is expanded as
{\small\begin{equation}
P_{out,GS}^{UB}\left(\rho_{th}\right)=\Pr\left(\underline{\rho}{}_{min,GS}\leq\rho_{th}\right)=\left[\Pr\left(\zeta_{min}^{\left(J'\right)}\leq\rho_{th}\right)\right]^{M},\label{eq:iid_UB}
\end{equation}}where $\zeta_{min}^{\left(J'\right)}$ is defined in (\ref{eq:approximation1}).
It is also noted that the approximation of $\Pr\left(\zeta_{min}^{\left(J'\right)}\leq\rho_{th}\right)$
can be easily obtained by substituting $M_{1}=M_{2}=M_{3}=1$ into
(\ref{eq:I1}), (\ref{eq:I2}), (\ref{eq:App1}), and after some
calculations, we have the result in (\ref{eq:UB1}).

\section*{Appendix V Derivations for the High SNR approximations }

The high SNR approximations aim to facilitate the extraction of the
MuD gain. Although the results in (\ref{eq:UB}) and (\ref{eq:UB1})
are easy to be evaluated with software, they involve integral of exponential
functions and are not easy to be further analyzed. To this end, some
approximations are applied before the high SNR analysis. Specifically,
$I_{1,3}^{'}$ in (\ref{eq:App1}) is further approximated as $I_{1,3}^{'}<I_{1,3}^{''}=\int_{0}^{\infty}\int_{0}^{\infty}\int_{0}^{a+\frac{b}{y}}f\left(x,y,z\right)dxdydz,$
where the integral intervals of $x$ and $y$ have been enlarged from
$x,\, y\in\left[\mu,\infty\right)$ to $x,\, y\in\left[0,\infty\right)$.
The rational of this approximation is that when SNR is high, the condition
$\mu\rightarrow0$ holds, and such approximation will not influence
the asymptotic behaviors that are relevant to the MuD gain. Then the
result of $I_{1,3}^{''}$ is given by
{\small
\begin{eqnarray}
I''_{1,3} & = & 2M_{2}\sum_{q=0}^{M_{2}-1}\left[\frac{\left(-1\right)^{q}\tbinom{M_{2}-1}{q}}{2\left(q+1\right)}+3\sum_{p=1}^{M_{1}}\tbinom{M_{2}-1}{q}\tbinom{M_{1}}{q}\left(-1\right)^{q+p}e^{-3pa}\sqrt{\frac{pb}{\left(q+1\right)}}K_{1}\left(6\sqrt{p\left(q+1\right)b}\right)\right]\nonumber \\
 &  & \,\,\,\,\,\,\,\times\left[1-\left(1-e^{-3\mu}\right)^{M_{3}}\right].\label{eq:App2}
\end{eqnarray}
}By using $I_{1,3}^{''}$,
$\Pr\left\{ E_{1}\left(\rho_{th}\right)\right\} $ is further approximated
as $\Pr\left\{ E_{1}\left(\rho_{th}\right)\right\} <\bar{\Pr}\left\{ E_{1}\left(\rho_{th}\right)\right\} =I_{1,1}+I_{1,2}+I_{1,3}^{''}$.
Following the same lines of the derivations of $P_{out,CS}^{UB}$,
we have the approximation $P_{out,CS}^{UB}<\bar{P}_{out,CS}^{UB}=\sum_{i=1}^{6}\bar{\Pr}\left\{ E_{i}\left(\rho_{th}\right)\right\} $.
It is noted that $\bar{P}_{out,CS}^{UB}$ can be obtained by replacing
the second item of summation in ({\ref{eq:UB}}) with a slight modification
of $I''_{1,3}$, i.e., changing $M_{k}$ into $M_{i,k}$, $k\in[1,3]$, which is given by
{\small{}
\begin{align}
\bar{P}_{out,CS}^{UB}\left(\rho_{th}\right) & \approx\sum_{i=1}^{6}\left(\frac{M_{i,2}M_{i,3}\left(1-e^{-3\mu}\right)^{M_{\Sigma}}+M_{i,2}M_{\Sigma}\left(1-e^{-3\mu}\right)^{M_{i,1}+M_{i,2}}\left[1-\left(1-e^{-3\mu}\right)^{M_{i,3}}\right]}{\left(M_{i,1}+M_{i,2}\right)M_{\Sigma}}\right.\nonumber\\
 & +2M_{i,2}\sum_{q=0}^{M_{i,2}-1}\left[\frac{\left(-1\right)^{q}\tbinom{M_{i,2}-1}{q}}{2\left(q+1\right)}+3\sum_{p=1}^{M_{i,1}}\tbinom{M_{i,2}-1}{q}\tbinom{M_{i,1}}{q}\left(-1\right)^{q+p}e^{-3pa}\sqrt{\frac{pb}{\left(q+1\right)}}K_{1}\left(6\sqrt{p\left(q+1\right)b}\right)\right]\nonumber \\
 & \left.\times\left[1-\left(1-e^{-3\mu}\right)^{M_{i,3}}\right]\right).\label{eq:UB122}
\end{align}}On the other hand, by substituting $M_{1}=M_{2}=M_{3}=1$
in $\bar{P}_{out,CS}^{UB}$ and following the derivation of $P_{out,GS}^{UB}$,
we have $P_{out,GS}^{UB}<\bar{P}_{out,GS}^{UB}$, and $\bar{P}_{out,GS}^{UB}$
is obtained by replacing the third item in (\ref{eq:UB1}) with $6e^{-3\mu}\left(1-e^{-3a}6\sqrt{b}K_{1}\left(6\sqrt{b}\right)\right)$, which is given by
{\small{}
\begin{equation}
P_{out,GS}^{UB}\left(\rho_{th}\right)\approx\left[3\left(1-e^{-3\mu}\right)^{2}-2\left(1-e^{-3\mu}\right)^{3}+6e^{-3\mu}\left(1-e^{-3a}6\sqrt{b}K_{1}\left(6\sqrt{b}\right)\right)\right]^{M}.\label{eq:UB111}
\end{equation}
} Then, $P_{out,C(G)}^{LB}$ and $\bar{P}_{out,CS(GS)}^{UB}$ are actually
used to continue the high SNR analysis.

For the high SNR analysis, we first approximate $b=\frac{6\rho_{th}\left(\mathrm{SNR}+1\right)}{\mathrm{SNR^{2}}}\approx\frac{6\rho_{th}}{\mathrm{SNR}}=3a$
and $\mu\approx\frac{1}{2}\left(a+\sqrt{a^{2}+12a}\right)$, consequently,
$P_{out,C(G)}^{LB}$ and $\bar{P}_{out,C(G)}^{UB}$ can be defined
as functions of $a$. Then the power series of the exponential function
$e^{x}$ and the modified Bessel function $K_{1}\left(x\right)$ \cite{jeffrey2007table}
are used to express $P_{out,C(G)}^{LB}$ and $\bar{P}_{out,C(G)}^{UB}$
with the polynomial forms regarding $a$. By finding the first nonzero
derivative orders of these polynomials and discarding the higher order
infinitesimal terms when $a\rightarrow0$, the key results are obtained
and shown in (\ref{eq:LB_High})-(\ref{eq:UB_High_G}).

\section*{Appendix VI Proof of Proposition 3}
\begin{IEEEproof}
Let us focus on the CS scenario first, the LB of $P_{out,CS}\left(\mathrm{SNR}\right)$
is given by
{\small\begin{eqnarray}
P_{out,CS}\left(\mathrm{SNR}\right) & \geq & \Pr\left(R_{CS}^{UB}\leq R_{th}\left(\mathrm{SNR}\right)\right)\nonumber \\
 & = & \Pr\left(\overline{\rho}_{min,CS}\leq\left(1+\mathrm{SNR}\right)^{\frac{r}{3}}-1\right)\nonumber \\
 & = & P_{out,CS}^{LB}\left(\mathrm{SNR}\right),\label{eq:Pout_LB}
\end{eqnarray}}where $R_{CS}^{UB}=3\log_{2}\left(1+\overline{\rho}_{min,CS}\right)$ is
the UB of $R$. Then, according to the high SNR analysis in (\ref{eq:LB_High})
and Lemma 5 in Appendix I, it is easy to check that
{\small\begin{equation}
P_{out,CS}^{LB}\dot{=}\mathrm{SNR}^{-d_{CS}^{UB}\left(1-\frac{r}{3}\right)}.\label{eq:dmt_UB}
\end{equation}}
 One the other hand, the UB of $P_{out,CS}\left(\mathrm{SNR}\right)$
is given by
{\small\begin{eqnarray}
P_{out,CS}\left(\mathrm{SNR}\right) & \leq & \Pr\left(R_{CS}^{LB}\leq R_{th}\left(\mathrm{SNR}\right)\right)\nonumber \\
 & = & \Pr\left(\underline{\rho}{}_{min,CS}\leq\left(1+\mathrm{SNR}\right)^{\frac{r}{3}}-1\right)\nonumber \\
 & = & P_{out,CS}^{UB}\left(\mathrm{SNR}\right),\label{eq:Pout_UB}
\end{eqnarray}}where $R_{CS}^{LB}=3\log_{2}\left(1+\underline{\rho}{}_{min,CS}\right)$
is the lower bound of $R$. Based on (\ref{eq:UB_High}) and Lemma
5 in Appendix I, it is shown that
{\small\begin{equation}
P_{out,CS}^{UB}\left(\mathrm{SNR}\right)\dot{=}\mathrm{SNR}^{-d_{CS}^{LB}\left(1-\frac{r}{3}\right)}.\label{eq:dmt_LB}
\end{equation}}Noting $d_{CS}^{UB}=d_{CS}^{LB}=\min\left\{ M_{1},M_{2},M_{3}\right\} ,$
the DMT in (\ref{eq:DMT1}) for CS scenario is obtained by combining
(\ref{eq:Pout_LB}) to (\ref{eq:dmt_LB}). The DMT analysis for the
GS scenario follows the similar procedures as the CS scenario, and
we omit this part for limited space. The proof is then finished.
\end{IEEEproof}

\section*{Appendix VII Complexity Analysis in Table II}

The focus of the brief complexity analysis is to highlight the distributed nature of the proposed schemes. In particular, the number of users ($M$) is of particular interest. In the following part of this appendix, we briefly explain the calculation results given in Table II.

The complexity at Relay of centralized CS with Min-UA/ER-UA: The relay
needs to 1) enumerate all $M^{3}$ possible user combinations according to the selection criteria in (\ref{eq:Min-UA-C}) and (\ref{eq:outage optimal}), and for each combination the rely
needs to 2) perform the SSA and other related signal processing. Since
the computational complexity of 2) can be treated as a constant given
specific antenna configuration at user/relay, we can omitted it in
the big-O notation if necessary. So, we can obtain the complexity
as $O\left(M^{3}\right)$.

The complexity at User of distributed CS/GS with Min-UA/ER-UA: It
is noted that each user only needs to calculate the simple scheduling
metric(s) with (\ref{eq:D_CS}) or (\ref{eq:d-UE}); therefore, the involved computation complexity can be treated
as a constant given specific antenna configuration at user. Then we
can have the complexity as $O\left(1\right)$.

The complexities at Relay of centralized and distributed GS with Min-UA/ER-UA:
Firstly, we note that the relay needs to 1) enumerate all $M$ groups according to the selection criteria in (\ref{eq:d_GS}) and (\ref{eq:group-wise distributed}),
and for each group the relay needs to 2) compute the scheduling metric.
Based on this observation, we can initially conclude that the complexities
could be $O(M)$ for both cases. However, in order to show the difference,
we should further specify the constants regarding 2) the computation
of scheduling metric. It is noted that for the centralized GS with Min-UA,
the relay needs to do a couple of relatively complicated matrix operations
to extract the scheduling metrics in (\ref{eq:new_post_processing_SNR}) and (\ref{eq:Min-UA-C}), where the complexity
is denoted $a_{1}$. In contract to the centralized scheme, the relay only needs to synthesize the feedbacks from users to get a scheduling metric with (\ref{eq:pre_angle})-(\ref{eq:d_GS}), where the complexity is denoted as $a_{2}$.
Therefore, we can show the complexities of centralized and distributed
GS with Min-UA as $O(a_{1}M)$ and $O(a_{2}M)$, where $a_{1}\gg a_{2}$.
Finally, following similar arguments with references to (\ref{eq:Link_Pout})
and (\ref{eq:group-wise centralized}), we can show the complexities of the centralized and distributed
GS with ER-UA as $O(b_{1}M)$ and $O(b_{2}M)$, where $b_{1}\gg b_{2}$.

}

 \bibliographystyle{IEEEtran}
\bibliography{symbolbasedpnc}

\end{document}